\documentclass{emulateapj}
\shorttitle{The SLUGGS Survey}
\shortauthors{Brodie \etal~}
\def\etal{{\it et al.}}
\def\kms{\,km~s$^{-1}$}

\def\Reff{$R_{\rm e}$}
\def\vrms{$v_{\rm rms}$}
\def\atlas3d{ATLAS$^{\rm 3D}$}

\def\lsim{\mathop{\hbox{${\lower3.8pt\hbox{$<$}}\atop{\raise0.2pt\hbox{$\sim$}}
$}}}
\def\gsim{\mathop{\hbox{${\lower3.8pt\hbox{$>$}}\atop{\raise0.2pt\hbox{$\sim$}}
$}}}
     
\usepackage{subfigure}
\usepackage{graphics}
\usepackage{natbib}
\usepackage{journal_names}

\begin{document}

\title{The SAGES Legacy Unifying Globulars and GalaxieS Survey (SLUGGS):\\ 
Sample definition, methods, and initial results}

\author{Jean P.~Brodie\altaffilmark{1}, Aaron J.~Romanowsky\altaffilmark{1,2}, Jay Strader\altaffilmark{3}, Duncan A.~Forbes\altaffilmark{4}, 
Caroline Foster\altaffilmark{5}, Zachary G. Jennings\altaffilmark{1}, Nicola Pastorello\altaffilmark{4}, Vincenzo Pota\altaffilmark{1,4}, Christopher Usher\altaffilmark{4}, 
Christina Blom\altaffilmark{4}, Justin Kader\altaffilmark{1}, Joel C. Roediger\altaffilmark{1}, Lee R. Spitler\altaffilmark{5,6}, 
Alexa Villaume\altaffilmark{1},  
Jacob A. Arnold\altaffilmark{1},  Sreeja S.~Kartha\altaffilmark{4}, Kristin A.~Woodley\altaffilmark{1}}

\affil{
\altaffilmark{1}University of California Observatories, 1156 High Street, Santa Cruz, CA 95064, USA; 
	{\tt jbrodie@ucsc.edu}\\
\altaffilmark{2}Department of Physics and Astronomy, San Jos\'e State University, One Washington Square, San Jose, CA 95192, USA\\
\altaffilmark{3}Department of Physics and Astronomy, Michigan State University, East Lansing, Michigan 48824, USA\\
\altaffilmark{4}Centre for Astrophysics \& Supercomputing, Swinburne University, Hawthorn, VIC 3122, Australia\\
\altaffilmark{5}Australian Astronomical Observatory, PO Box 915, North Ryde, NSW 1670, Australia\\
\altaffilmark{6}Department of Physics and Astronomy, Macquarie University, North Ryde, NSW 2109, Australia
}

\slugcomment{The Astrophysical Journal, in press}

\begin{abstract}

We introduce and provide the scientific motivation for a wide-field photometric and spectroscopic chemodynamical survey of nearby early-type galaxies (ETGs) and their globular cluster (GC) systems. The SLUGGS\footnote{\tt http://sluggs.ucolick.org} (SAGES Legacy Unifying Globulars and GalaxieS) survey is being carried out primarily with Subaru/Suprime-Cam and Keck/DEIMOS. The former provides deep $gri$ imaging over a 900 arcmin$^{2}$ field-of-view to characterize GC and host galaxy colors and spatial distributions, and to identify spectroscopic targets. The NIR \ion{Ca}{2} triplet
provides GC line-of-sight velocities and metallicities out to typically $\sim$\,8\,\Reff, and to $\sim$\,15\,\Reff\ in some cases.  New techniques to extract integrated stellar kinematics and metallicities to large radii ($\sim$\,2--3\,\Reff) are used in concert with GC data to create two-dimensional velocity and metallicity maps for comparison with simulations of galaxy formation. 
The advantages of SLUGGS compared with other, complementary, 2D-chemodynamical surveys are its superior velocity resolution, radial extent, and
multiple halo tracers. We describe the sample of 25 nearby ETGs, the selection criteria for galaxies and GCs, the observing strategies, the data reduction techniques, and modeling methods.
The survey observations are nearly complete and more than 30 papers have so far been published using SLUGGS data. Here we summarize some initial results, including signatures of two-phase galaxy assembly, evidence for GC metallicity bimodality, and a novel framework for the formation of extended star clusters and ultracompact dwarfs.  An integrated overview of current chemodynamical constraints on GC systems points to separate, in-situ formation modes at high redshifts for metal-poor and metal-rich GCs.

\end{abstract}


\section{Introduction and Scientific Motivation}\label{sec:intro}

Recent advances in both observations and theory of early-type galaxies (ETGs; ellipticals and lenticulars)
have begun to revolutionize our understanding of these systems as more than simply ``red and dead''
(where star formation has long since ceased).  
High-redshift ($z$) studies have revealed an unexpectedly strong systematic size evolution,where galaxies in the early universe were much denser and more compact than their 
present-day counterparts \citep[e.g.,][]{Daddi05,vanDokkum09}. 
These observations are now widely understood on theoretical grounds to reflect 
a ``two-phase'' or ``inside-out'' formation history, where an early, rapid phase of assembly and star formation establishes the cores
of galaxies  (on $\sim$ 1 kpc scales), followed by a protracted expansion of the galaxy through minor mergers 
(as opposed to a simple major-merger scenario; 
\citealt{Khochfar06, Naab09, Hopkins10a, Oser10, Oser12, Johansson12, Lackner12, Hilz12, Hilz13, Gabor12, Oogi13, Bedorf13,Porter14a};
but see \citealt{Nipoti12}).

\begin{figure}
\includegraphics[width=\columnwidth]{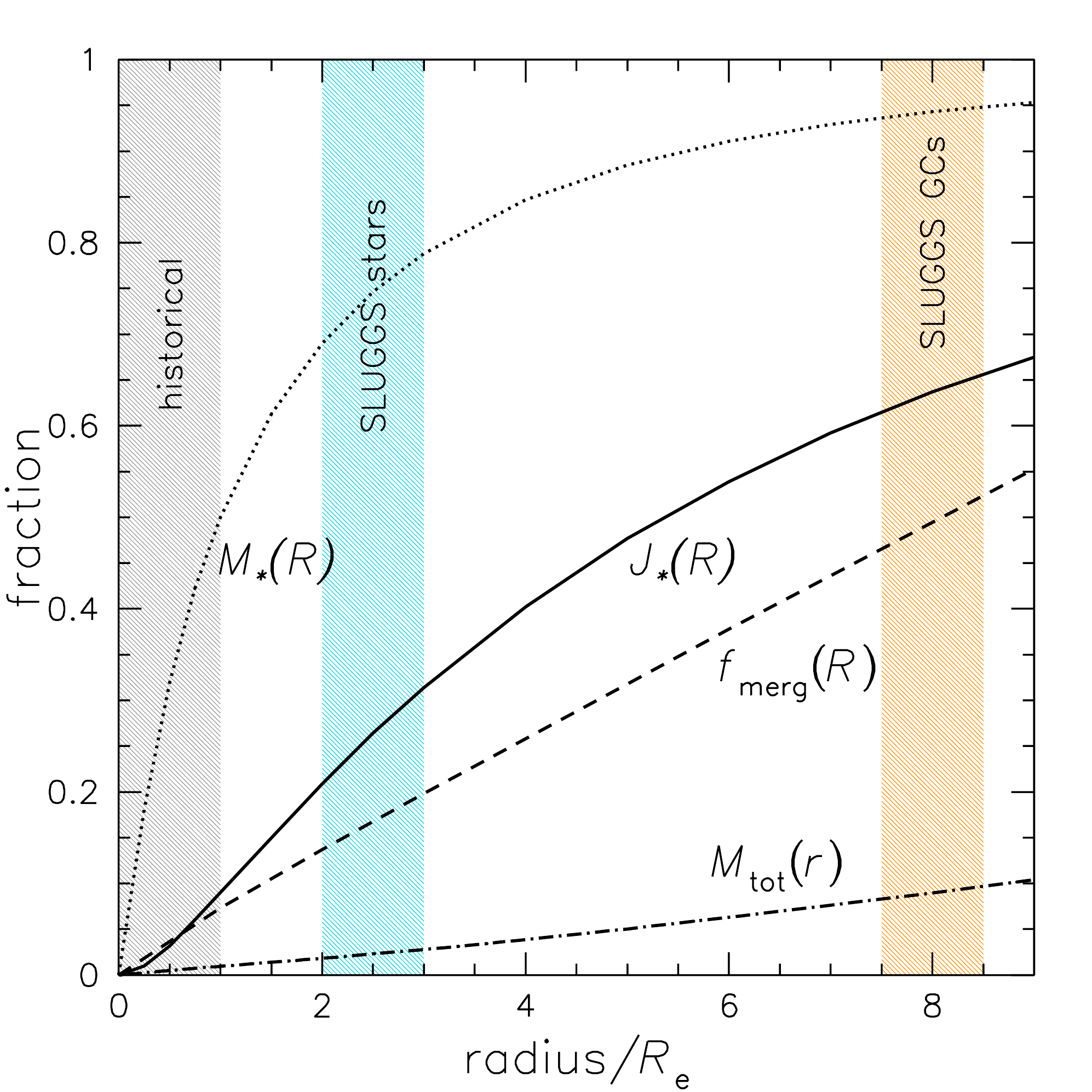}
\caption{Cumulative fraction of physical quantities versus radius in an ordinary elliptical galaxy. 
The galaxy is approximated as a very simple $R^{1/4}$ projected density profile with a rotation velocity that is constant with radius.  
The projected stellar mass and angular momentum profiles are thus shown as $M_*(R)$  and $J_*(R)$.  
The total mass profile including dark matter, $M_{\rm tot}(r)$, and the probability of detectable substructure at a given
radius, $f_{\rm merg}(R)$, are derived from cosmologically-based models (Sections~\ref{sec:mass} and \ref{sec:twophase}).
 The shaded area at left reaches out to one effective radius,  the typical extent of detailed galaxy kinematics studies to date, 
 which therefore miss $\sim$\,50\% of $M_*$, $\sim$\,90\% of $J_*$,  $\sim$\,99\% of $M_{\rm tot}$,
 and $\sim$\,93\% of the potential merging signatures.
 The other shaded areas show the radii reached by SLUGGS -- capturing much larger fractions of these important parameters.
 }
 \label{fig:fracs}
\end{figure}

This picture may be tested and refined further through
observations of low-$z$ galaxies, whose outer regions are natural proving grounds for the later accretion phase of galaxy assembly, owing to the long dynamical timescales
(Figure~\ref{fig:fracs}) and more recent putative activity in these regions.

Various studies have pursued this idea by searching for transitions and substructures
in the stellar halos of nearby ETGs \citep[e.g.,][]{Forbes92,Mihos05,Mihos13,Harris07,Tal09,Coccato10,Mouhcine11,LaBarbera12,Huang13b,Greene13,DSouza14}.

In this context, what has been missing is a dedicated, systematic survey of the wide-field chemodynamics of nearby ETGs, 
akin to the SAURON and \atlas3d\ surveys \citep{deZeeuw02,Cappellari11a} but on much larger spatial scales.
These surveys, like the vast majority of ETG studies, have focused on the central regions, well inside an 
effective radius \Reff\ that encloses half of a galaxy's projected stellar mass or luminosity.
By definition, this misses over half of the stars, including virtually all those in the crucial halo regions.

More acutely, dynamical studies of halos are required to constrain 
two of the most fundamental parameters that characterize a galaxy:
the stellar angular momentum and the total mass (including dark matter), 
which are typically only $\sim$ 10\% and $\sim$ 5\% contained inside 1~\Reff, respectively (Figure ~\ref{fig:fracs}). 
The halo mass is a key controlling parameter in galaxy formation models. However,
connections between ordinary ETGs and their halo masses are 
usually made only on a statistical (rather than individual) basis, and often through indirect means \citep[e.g.,][]{Yang05}. 
The angular momentum traces torques in galaxy formation and is an important benchmark for theoretical models
 \citep[e.g.,][]{Mestel63,Fall80,Navarro97a,Governato04,Obreschkow14}, but estimates of its total value in ETGs 
have so far been based on a hodge-podge of inhomogeneous
data from the literature \citep{RomanowskyFall12}.

Pushing spectroscopy out into the halos of ETGs is very challenging because of the low surface brightness levels of the stellar light,
with maximum galactocentric radii of $\sim$~4~\Reff\ currently attainable. Beyond these radii, it is necessary to use a bright discrete
tracer population. Planetary nebulae (PNe) are one useful option for kinematical constraints,
(e.g., \citealt{Douglas02,Peng04a,Coccato09,McNeil10,Ventimiglia11})
but using them to probe stellar populations is observationally expensive (e.g., \citealt{Mendez05}).

Globular clusters (GCs) have a distinguished history as indicators of the assembly histories of
galaxy halos, for the Milky Way and beyond \citep{Searle78, Brodie91,Abadi06,Brodie06,Forbes10}.
They provided evidence that ETG evolution was driven by minor rather than major mergers, 
long before this became the mainstream model \citep{Forte82,Cote98,Kissler-Patig98a}.
They have also long been used to infer the presence and properties of  metal-poor stellar halos,
which are a subject of intense interest in galaxy formation (e.g., \citealt{Bullock05,Cooper10})
but which have been very challenging to study directly even in the nearest galaxies
(e.g., \citealt{Kalirai06,Harris07,Mihos13,Ibata14}).

Many insightful imaging studies of GCs have been carried out 
using the {\it Hubble Space Telescope} ({\it HST}, which generally covers only the central regions of galaxies; e.g., \citealt{Larsen01, Kundu01, Peng06, Strader06, Mieske10}),
and from the ground (which affords a much wider field of view; e.g., \citealt{Forte07, Hargis12}). However,
there has so far not been a homogeneous {\it spectroscopic} survey of GC systems for a large sample of ETGs covering
the full range of environments.

\subsection{SLUGGS}

Into this void, we have launched the SAGES Legacy Unifying Globulars and GalaxieS (SLUGGS) survey.
The survey uses a combination of imaging from Subaru/Suprime-Cam and spectroscopy from Keck/DEIMOS to obtain the {\it spatial},
{\it kinematical}, and {\it chemical} properties of both  {\it diffuse stellar light} and {\it GCs} in two dimensions and
over wide areas around 25 nearby ETGs spanning a broad range of galaxy properties and environments. 

Some aspects of the survey, such as the GC imaging and spectroscopy, build on established observational techniques and
apply them with great success using state-of-the-art instrumentation.
Other aspects are novel and potentially transformative for the study of nearby galaxies.
These  involve Stellar Kinematics from Multiple Slits (SKiMS),
a technique that we have pioneered to map out the kinematics and metallicities of diffuse starlight in 2D to large radii
(e.g., \citealt{Norris08,Proctor08,Proctor09}).

All of the spectroscopic aspects of the survey exploit the strong \ion{Ca}{2} triplet absorption line feature
in the near-infrared ($\sim$\,8500\,\AA), where DEIMOS has both a high efficiency and a high spectral resolution for separating
the galaxy and GC lines from the forest of near-infrared sky lines.

The main advantages of the SLUGGS survey are the superior velocity precision (median 12\,\kms) 
on a very stable, high-throughput instrument, and the large radial coverage (typically $\sim$\,2--3\,\Reff\ for stars,  
extending to $\sim$\,8\,\Reff\ when GCs are included).

\begin{figure}
\includegraphics[width=\columnwidth]{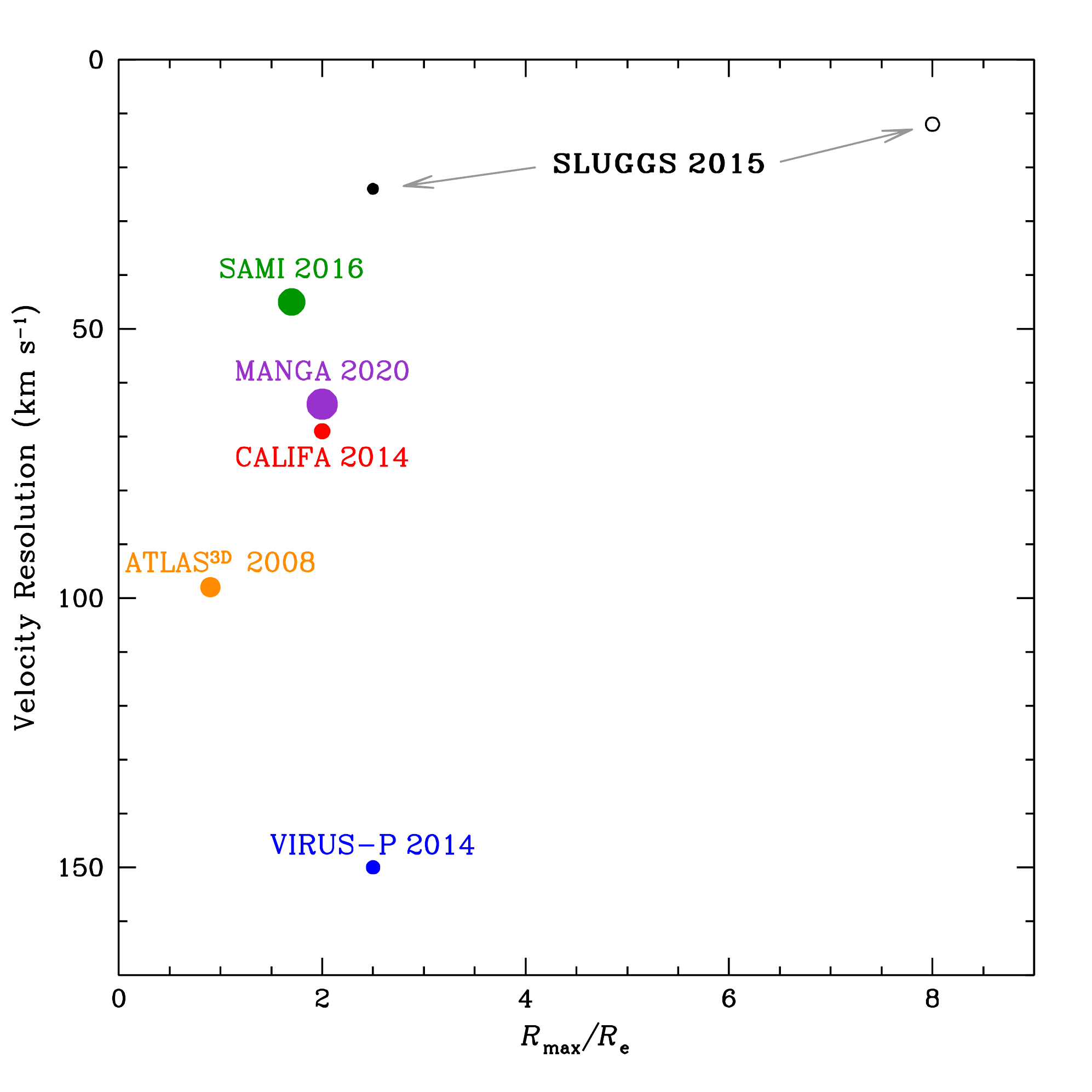}
\caption{SLUGGS survey figure of merit. Instrumental velocity resolution ($\sigma$) is 
plotted (with superior resolution at the top) vs.\  maximum galactocentric radial extent for a typical survey galaxy, 
in units of effective radii. The symbols  show the different 2D chemodynamical surveys with their data completion 
dates indicated in the legend. Symbols are scaled by the log of the number of 
early-type galaxies surveyed. The open circle represents the globular cluster aspect of the SLUGGS survey.
For the MaNGA survey we have combined the primary and secondary samples.
SLUGGS is complementary to previous, or currently underway,  
2D spectroscopic surveys; it covers a more limited number of galaxies but offers better radial coverage and higher velocity  resolution.
}
\label{fig:merit}
\end{figure}

\begin{figure}
\centering
\includegraphics[angle=0,width=\columnwidth]{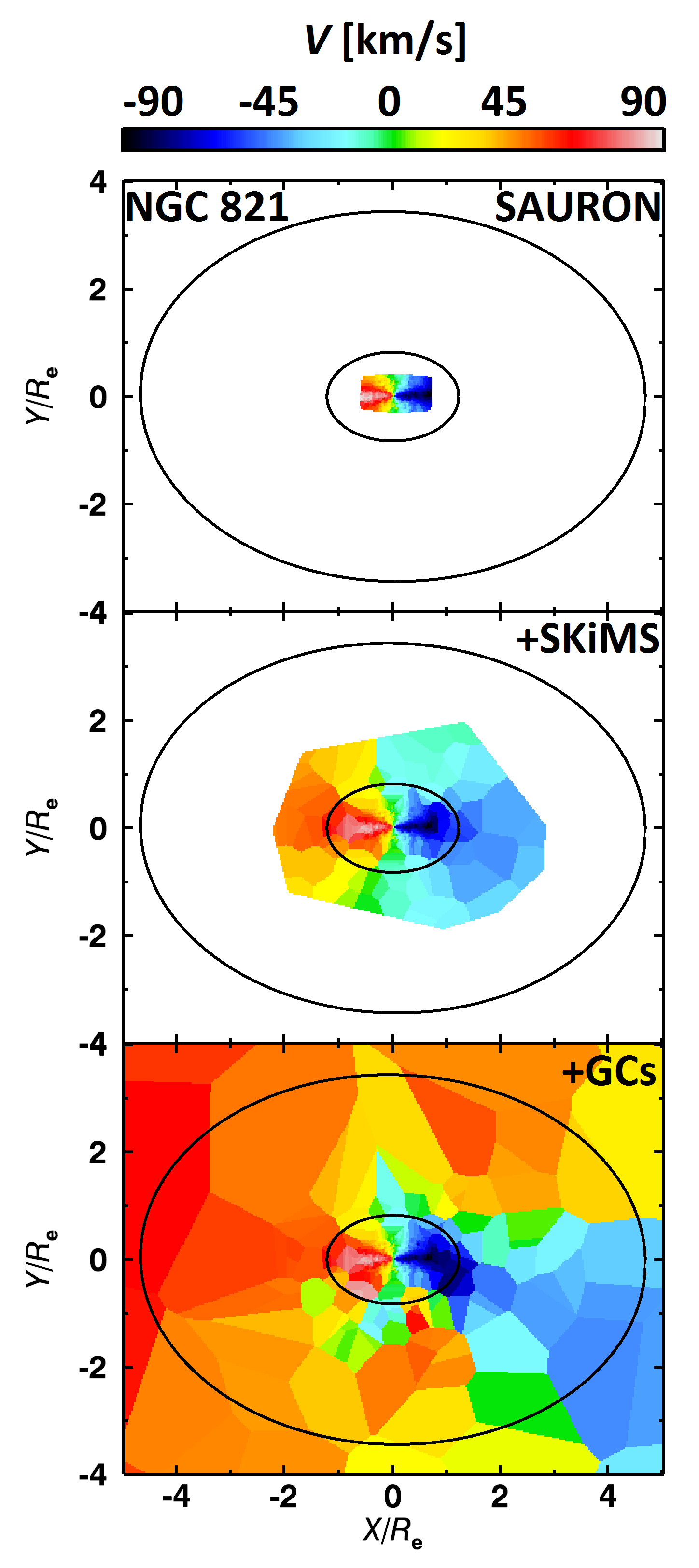}
\caption{
Stellar mean line-of-sight velocity field of the elliptical galaxy NGC~821, where the color scale indicates the velocity amplitude
(see scale-bar at top). Ellipses mark schematic isophotes at $\sim$\,1\,\Reff\ and $\sim$\,4\,\Reff.
The top panel shows an \atlas3d/SAURON pointing of the central region \citep{Cappellari11a},
the middle panel incorporates the intermediate-radius regions 
reconstructed from SLUGGS/DEIMOS data using the SKiMS (stellar kinematics with multiple slits) technique
along with major- and minor-axis long-slit data from the literature, and the bottom panel folds in GC velocities to large radii.
Note that the SLUGGS methods have sparser spatial sampling than a true integral-field unit, requiring some smoothing
in creating these maps (described in Romanowsky et al., in preparation).
The central regions do not reveal the kinematic twisting and complexity
that are apparent in the wide field map.
}
\label{fig:layer}
\end{figure}

\subsection{Survey Comparisons}\label{sec:survcomp}

\begin{table*} 
\caption{2D Chemo-dynamical Surveys}\label{surveys}
				\begin{tabular*}{18cm}{@{\extracolsep{\fill}}@{}cccccccc@{}}\toprule
				\multicolumn{0}{c}{Survey} &
				\multicolumn{0}{c}{SLUGGS} &
				\multicolumn{0}{c}{ATLAS$^{\rm 3D}$} &
				\multicolumn{0}{c}{VIRUS--P} &
				\multicolumn{0}{c}{MASSIVE} & 
				\multicolumn{0}{c}{CALIFA} &
				\multicolumn{0}{c}{SAMI} &
				\multicolumn{0}{c}{MaNGA} \\
\hline
Completion Year & 2015 & 2008 & 2014 & $>2014$ & 2015 & 2016 & 2020\\
No. of ETGs & 25 & 260 & 50 & 116 &  80 & 1700 & 3500, 1500\\
Distance (Mpc) & $<30$ & $<42$ & 40--95 & 17--108 & 20--120 & $\sim$~170 & $\sim$~130\\
Radial extent ($R_{\rm e}$) & 2.5, 8 & 0.9 & 2.5 & 2.1 & 2 & 1.7 & 1.5, 2.5\\
Vel. Res. $\sigma$ (\kms) & 24 & 98 & 150 & 150 & 69 & 73 & 64\\
Wavelengths (\AA) & 6500--9100 & 4800--5380 & 3550--5850 & 3550--5850 & 3700--5000 & 3700--5700 & 3600--10,000\\   
\hline 
\end{tabular*} 
\\ 
Rows give survey or instrument name, expected year of completion of data acquisition, number 
of early-type galaxies (ETGs) studied (note that there are two different radial extents in SLUGGS for
stars and GCs, and in MaNGA for the primary and secondary samples),
typical distance of sample, typical radial extent probed,  instrumental velocity resolution $\sigma$ (note that SAMI
also has higher resolution coverage at redder wavelengths that is used for emission-line mapping rather than
integrated stellar spectroscopy), total wavelength range. 
\end{table*}

In Figure~\ref{fig:merit} and Table~\ref{surveys} we have summarized other current and planned 2D chemodynamical surveys of ETGs.
These surveys use either an Integral Field Unit (IFU) or a fiber-bundle. We briefly discuss these in turn.

Building on the success of the original SAURON
survey, the \atlas3d\ survey \citep{Cappellari11a} of 260 nearby ETGs has had a high scientific
impact. However, it has a moderate velocity resolution of 105\,\kms\ 
(limiting its ability to map both low-mass galaxies and kinematically cold features in all galaxies)
and, perhaps most importantly, a small field-of-view. To compensate for this, galaxies
were often targeted with two pointings, giving a median extent for the survey observations of 0.9\,\Reff. For more luminous 
and nearby galaxies, comparable to those in SLUGGS, the typical extent is only 0.6\,\Reff.

A recently initiated survey comprising all galaxy types using the SAMI instrument \citep{Bryant14}
will increase the number of ETGs mapped with higher velocity resolution and wider wavelength coverage, but
will not reach much beyond 1~\Reff.  As noted earlier, most of the mass and angular momentum in ETGs is located beyond 1\,\Reff,
where we also expect key signatures of galaxy formation.

A number of surveys are probing beyond 1\,\Reff. CALIFA \citep{Sanchez12}
is well underway and, upon completion (expected in 2015), around 80 
ETGs will have been observed out to about 2\,\Reff\ with a velocity resolution of 69\,\kms\ (blue arm).

Perhaps most comparable to SLUGGS is the study of 50 nearby ETGs that uses the Mitchell Spectrograph IFU (formerly VIRUS-P;
\citealt{Greene13,Raskutti14}). 
This study operates in the optical and so is capable of obtaining both age and metallicity information, as well as kinematics, out
to $\sim$~2.5~\Reff. 
The drawbacks are the relatively poor velocity resolution (150~\kms) and the lower S/N at large radii, which inhibits the 2D coverage.
A follow-up survey, MASSIVE, is focusing on galaxies at the extreme high end of the stellar mass range \citep{Ma14}.

Due to start in 2014 and collect data for six years, the MaNGA survey 
(PI Kevin Bundy, {\tt http://www.sdss3.org/future/manga.php})  will map an impressive 10,000 galaxies, roughly half of which will be
ETGs. It will cover UV to near-IR wavelengths (3600--10,000 \AA) 
with a velocity resolution of 64 \kms. Although it will 
reach only about 1.5~\Reff\ for the primary sample,  a secondary sample of $\sim$1500 will be observed to 2.5~\Reff. 

Partially overlapping with SLUGGS both in theme and in galaxy sample is
the Next Generation Virgo Cluster Survey (NGVS), a multi-band photometric study of the Virgo cluster with an emphasis on GCs
\citep{Ferrarese12}.
However, NGVS does not incorporate a full, systematic spectroscopic component, which is the main focus of our survey,
along with an exploration of galaxies in a wide range of environments.

\subsection{This Paper}

The SLUGGS survey has been designed to address the following science themes:
What are the basic, global chemodynamical properties of ETGs? (including their masses, angular momenta, orbital structures, and metallicities over a wide range in radius).
What is the distribution of dark matter in ETGs?
How are the outer regions of ETGs assembled?
How does ETG assembly depend on mass, environment, and other variables? 
Do the observations agree with theoretical models of galaxy formation?

SLUGGS began as a pilot project in 2006 November, using DEIMOS to measure velocities of GCs around the giant elliptical NGC~1407 
\citep{Romanowsky09}.  Since then it has evolved into a full survey, leading to over 30 papers to date (many of them discussed below),
with the first large installments of data coming in \citet{Usher12} on GC metallicities, \citet{Pota13a} on GC kinematics,
\citet{Arnold14} on stellar kinematics, and \citet{Pastorello14} on stellar metallicities.

As a preview of the nature and quality of the SLUGGS results, 
Figure~\ref{fig:layer} shows the observed 2D stellar velocity field
of a galaxy, zooming out to show different types of data on three spatial scales.
The inner regions of the data come from the well-known SAURON
integral field spectrograph, the intermediate regions are from SKiMS, and the outer regions come from GCs.
The SKiMS data show initially the same pattern as SAURON out to 1\,\Reff, with a disklike rotation pattern that is
well-aligned with the isophotes. However, by 2\,\Reff, there are indications of a kinematic twist that becomes
even more prominent in the GC data at $\sim$\,5\,\Reff\ where there
are additional asymmetries that suggest  unrelaxed post-merger material.
Thus at large radii, we see the emergence of new information about the galaxy structure and formation
that is not apparent from the central regions alone.

This paper provides a general overview of SLUGGS, including the galaxy sample definition (Section~\ref{sec:samp}) and
summaries of our observing and data reduction techniques (Section~\ref{sec:obs}), and modeling methods (Section  \ref{sec:model}).
Some initial results are presented in Section~\ref{sec:res} and conclusions are provided in Section~\ref{sec:conc}.
An Appendix provides further details about the galaxy sample.
More information and news  can be found on the survey website\footnote{\tt http://sluggs.ucolick.org}.

\section{Galaxy Sample Selection and Properties}\label{sec:samp}

\begin{figure*}
\centering{
\includegraphics[width=18.1cm]{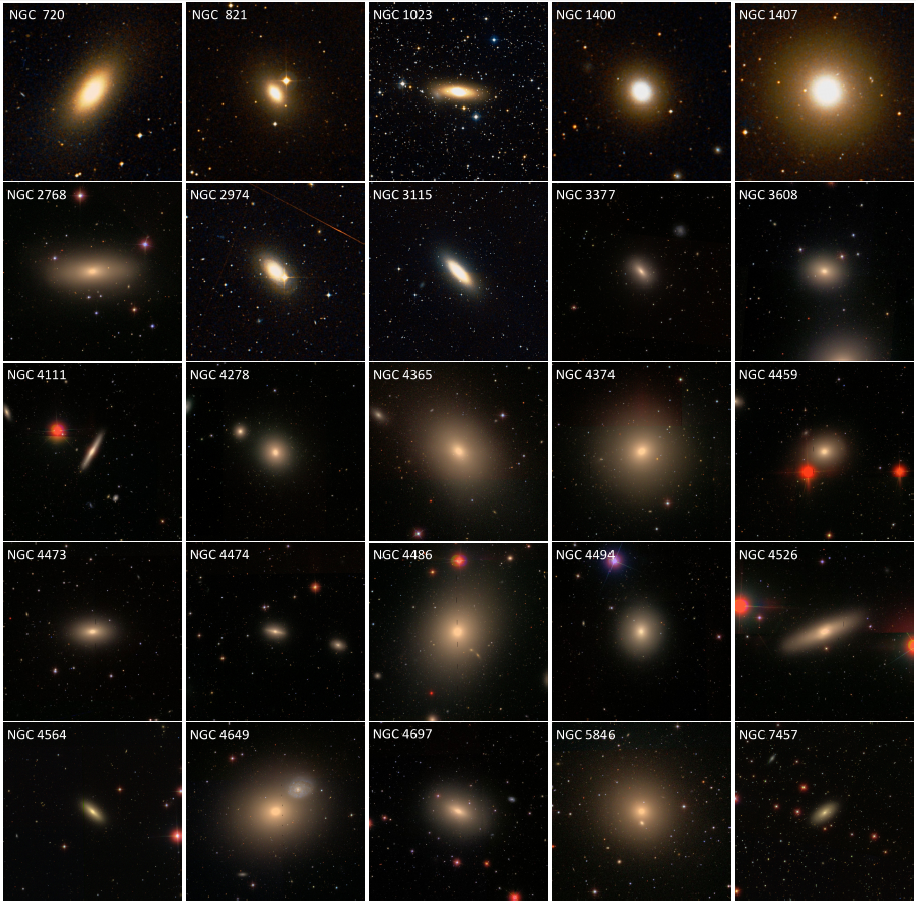} 
}
\caption{
Color thumbnails of the SLUGGS survey sample of galaxies, from the SDSS where available, and otherwise from the DSS
(provided via {\tt http://www.wikisky.org}).The field-of-view in each case is 70$\times$70~kpc.
North is up and East to the left. 
The galaxy identifications are as labeled in the panels.
 }\label{fig:thumbs}
\end{figure*}

Even with the best current observing resources, a proper volume- and 
magnitude-limited spectroscopic survey (like \atlas3d) is unfeasible for a 
study of GCs, so following the original SAURON survey \citep{deZeeuw02}, 
we explore a ``representative'' sample of 25 ETGs
(Figure~\ref{fig:thumbs} and Table~\ref{info}).
The strategy here is to consider the relevant basic parameters of these 
galaxies, and to identify the controlling factors for their formation histories 
and present day properties. Despite the limited pool of galaxies in the 
nearby universe (within $\sim$~25 Mpc), our sample has been constructed to provide 
excellent coverage of parameter space and to embrace the regime of 
``ordinary'' ETGs of the characteristic $L^*$ luminosity. These
galaxies in particular have so far been the subject of very little GC spectroscopy.

\begin{table*} 
\caption{}\label{info}
 \begin{tabular*}{18cm}{@{\extracolsep{\fill}}@{}rcrrcclcrcrrrcc@{}}\toprule
  \multicolumn{0}{c}{Galaxy} &
  \multicolumn{0}{c}{$M_K$} &
  \multicolumn{0}{c}{Dist.} &
  \multicolumn{0}{c}{$\sigma_{\rm kpc}$} &
  \multicolumn{0}{c}{$V/\sigma_{\rm e/2}$} &
  \multicolumn{0}{c}{\Reff} &
  \multicolumn{0}{c}{Morph.} &
  \multicolumn{0}{c}{$T_{\rm Hub}$} &
  \multicolumn{0}{c}{P.A.} &
  \multicolumn{0}{c}{$\epsilon$} &
  \multicolumn{0}{c}{$a_4/a$} &
  \multicolumn{0}{c}{$\gamma^{\prime}$} &
  \multicolumn{0}{c}{$V_{\rm sys}$} &
  \multicolumn{0}{c}{Env.} &
  \multicolumn{0}{c}{$\rho_{\rm env}$} \\
  \multicolumn{0}{c}{NGC} &
  \multicolumn{0}{c}{(mag)} &
  \multicolumn{0}{c}{(Mpc)} &
  \multicolumn{0}{c}{(km\,s$^{-1}$)} &
  \multicolumn{0}{c}{} &
  \multicolumn{0}{c}{($^{\prime\prime}$)} &
  \multicolumn{0}{c}{} &
  \multicolumn{0}{c}{} &
  \multicolumn{0}{c}{(deg)} &
  \multicolumn{0}{c}{} &
  \multicolumn{0}{c}{(\%)} &
  \multicolumn{0}{c}{} &
  \multicolumn{0}{c}{(km\,s$^{-1}$)} &
  \multicolumn{0}{c}{} &
  \multicolumn{0}{c}{(Mpc$^{-3}$)} \\
  \multicolumn{0}{c}{(1)}  &
  \multicolumn{0}{c}{(2)}  &
  \multicolumn{0}{c}{(3)}  &
  \multicolumn{0}{c}{(4)}  &
  \multicolumn{0}{c}{(5)}  &
  \multicolumn{0}{c}{(6)}  &
  \multicolumn{0}{c}{(7)}  &
  \multicolumn{0}{c}{(8)}  &
  \multicolumn{0}{c}{(9)}  &
  \multicolumn{0}{c}{(10)} &
  \multicolumn{0}{c}{(11)} &
  \multicolumn{0}{c}{(12)} &
  \multicolumn{0}{c}{(13)} &
  \multicolumn{0}{c}{(14)} &
  \multicolumn{0}{c}{(15)} \\
\hline
Main:&    &      &     &      &    &         &        &       &      &         &         &      &   &         \\
 720 & $-$24.88 & 26.9 & 227 & 0.10 & 35 & E5      & $-$4.9 & 140.0 & 0.49 &    0.00 &    0.07 & 1745 & F & 0.25 \\
 821 & $-$23.99 & 23.4 & 193 & 0.29 & 40 & E6      & $-$4.8 &  31.2 & 0.35 &    1.43 &    0.51 & 1718 & F & 0.08 \\
1023 & $-$24.01 & 11.1 & 183 & 0.37 & 48 & S0      & $-$2.6 &  83.3 & 0.63 &    0.68 &    0.74 &  602 & G & 0.57 \\
1400 & $-$24.35 & 26.8 & 236 & 0.23 & 28 & E1/S0   & $-$3.7 &  40.0 & 0.13 & $-$0.50 &    0.20 &  558 & G & 0.42 \\
1407 & $-$25.46 & 26.8 & 252 & 0.07 & 63 & E0      & $-$4.5 &  35.0 & 0.07 & $-$0.50 &    0.14 & 1779 & G & 0.42 \\
2768 & $-$24.71 & 21.8 & 206 & 0.24 & 63 & E6/S0   & $-$4.5 &  91.6 & 0.57 & $-$0.68 &    0.37 & 1353 & G & 0.31 \\
2974 & $-$23.62 & 20.9 & 231 & 0.81 & 38 & E4/S0     & $-$2.1 &  44.2 & 0.37 &    1.22 &    0.62 & 1887 & F & 0.26 \\
3115 & $-$24.00 &  9.4 & 248 & 0.59 & 35 & S0      & $-$2.9 &  40.0 & 0.66 &    4.00 &    0.52 &  663 & F & 0.08 \\
3377 & $-$22.76 & 10.9 & 135 & 0.56 & 36 & E5--6    & $-$4.8 &  46.3 & 0.33 &    0.74 &    0.68 &  690 & G & 0.49 \\
3608 & $-$23.65 & 22.3 & 179 & 0.05 & 30 & E1--2    & $-$4.8 &  82.0 & 0.20 & $-$0.34 &    0.17 & 1226 & G & 0.56 \\
4111 & $-$23.27 & 14.6 & 161 & 0.66 & 12 & S0      & $-$1.3 & 150.3 & 0.79 &    4.80 &       -- &  792 & G &   1.09 \\
4278 & $-$23.79 & 15.6 & 228 & 0.21 & 32 & E1--2    & $-$4.8 &  39.5 & 0.09 & $-$0.34 &    0.10 &  620 & G &   1.25 \\
4365 & $-$25.19 & 23.1 & 253 & 0.11 & 53 & E3      & $-$4.8 &  40.9 & 0.24 & $-$0.80 &    0.00 & 1243 & G &   2.93 \\
4374 & $-$25.13 & 18.5 & 284 & 0.03 & 53 & E1      & $-$4.3 & 128.8 & 0.05 & $-$0.27 &    0.25 & 1017 & C &   3.99 \\
4459 & $-$23.89 & 16.0 & 170 & 0.45 & 36 & S0      & $-$1.4 & 105.3 & 0.21 & $-$0.14 &    0.49 & 1192 & C &   4.06 \\
4473 & $-$23.76 & 15.2 & 189 & 0.26 & 27 & E5      & $-$4.7 &  92.2 & 0.43 &    0.95 &    0.10 & 2260 & C &   2.17 \\
4474 & $-$22.27 & 15.5 &  88 & 0.35 & 17 & S0      & $-$1.9 & 121.2 & 0.42 &    6.26 &    0.57 & 1152 & C &   3.80 \\
4486 & $-$25.31 & 16.7 & 307 & 0.02 & 81 & E0/cD   & $-$4.3 & 151.3 & 0.16 &    0.13 &    0.23 & 1284 & C &   4.17 \\
4494 & $-$24.11 & 16.6 & 157 & 0.22 & 49 & E1--2    & $-$4.8 & 176.3 & 0.14 &    0.03 &    0.55 & 1342 & G &   1.04 \\
4526 & $-$24.61 & 16.4 & 233 & 0.56 & 45 & S0      & $-$1.9 & 113.7 & 0.76 &    0.10 &       -- &  617 & C &   2.45 \\
4564 & $-$23.08 & 15.9 & 153 & 0.53 & 20 & E6      & $-$4.8 &  48.5 & 0.53 &    1.88 &    0.81 & 1155 & C &   4.09 \\
4649 & $-$25.36 & 16.5 & 308 & 0.12 & 66 & E2/S0   & $-$4.6 &  91.3 & 0.16 & $-$0.47 &    0.14 & 1110 & C &   3.49 \\
4697 & $-$24.13 & 12.5 & 180 & 0.36 & 62 & E6      & $-$4.5 &  67.2 & 0.32 &    1.30 &    0.82 & 1252 & G & 0.60 \\
5846 & $-$25.00 & 24.2 & 231 & 0.04 & 59 & E0--1/S0 & $-$4.7 &  53.3 & 0.08 & $-$0.34 & $-$0.02 & 1712 & G & 0.84 \\
7457 & $-$22.38 & 12.9 &  74 & 0.48 & 36 & S0      & $-$2.7 & 124.8 & 0.47 &    0.58 &    0.61 &  844 & F & 0.13 \\
\hline                                                                                         
Bonus: &   &      &     &      &    &         &        &       &      &         &         &      &   &         \\
3607 & $-$24.75 & 22.2 & 229 & 0.29 & 39 & S0      & $-$3.2 & 124.8 & 0.13 & $-$0.12 &    0.26 &  942 & G & 0.34 \\
4594 & $-$24.95 &  9.5 & 225 & 0.57 & 70 & Sa      &   +1.1 &  90.0 & 0.59 &       -- &       -- & 1024 & F & 0.32 \\
5866 & $-$24.00 & 14.9 & 163 & 0.35 & 36 & S0      & $-$1.2 & 125.0 & 0.58 &    6.19 &       -- &  755 & G & 0.24 \\
\hline 

\end{tabular*} 
\\ 
Parameters of the 25 galaxies from the SLUGGS Survey, as well as three bonus galaxies (at the bottom of the table).
Details of the parameter derivations are given in Section~\ref{sec:samp} and in the Appendix.
Column descriptions: (1) galaxy NGC number;
(2) extinction-corrected $K$-band absolute magnitude;
(3) distance measured in Mpc;
(4) central stellar velocity dispersion within 1\,kpc;
(5) rotational dominance parameter within 0.5\,\Reff;
(6) effective (half-light) radius; 
 (7) morphological type;
 (8) Hubble stage parameter, where $T_{\rm Hub}=-5$ for E, $-3$ for E/S0, $-2$ for S0, and $0$ for S0/a;
 (9) position angle and (10) ellipticity, for the outer isophotes;  
(11) isophote shape parameter;
(12) central stellar density slope;
(13) systemic velocity;
(14) local environment type: field (F), group (G), or cluster (C);
(15) environmental density of neighboring galaxies.
Alternative names for some of the galaxies are M84 (NGC~4374), M87 (NGC~4486), M60 (NGC~4649), and M104 (NGC~4594).
\end{table*}
Our selection parameters are described below.
They are all based on broad properties of the visible and near-infrared galactic stellar light, and are chosen for their
probable importance in galactic evolution, as well as for their general availability in the literature for nearby galaxies.
Where possible, we adopt the same parameters used in the \atlas3d\ survey, to provide a clear context
for how the SLUGGS sample fits in with the general population of galaxies in the nearby universe.
For the sake of homogeneity, the parameter values
are {\it not} always the most accurate that are available, as will be discussed further.

The first galaxy parameter is the $K$-band absolute magnitude, $M_K$, which for ETGs
is a very good proxy for stellar mass at the $\sim$\,10\% level (e.g., \citealt{Fall13}).
These values are based on 2MASS \citep{Jarrett00}, with further details provided in the Appendix.
Here it should be noted that the 2MASS photometry is known to systematically underestimate galaxy 
luminosities by $\sim$\,10--40\%, particularly for the most nearby, extended galaxies,
due to over-subtracting the sky background
(e.g., \citealt{Devereux09,Scott13}). 
Although this effect can be corrected for (e.g., \citealt{Romanowsky12}),
it is best done by using homogeneous model fits to the galaxy light profiles, which are
not yet available for our sample. 
Nonetheless, based on the scatter around the correction derived by \citet{Scott13}, $K_{\rm new} = 1.07 \times K_{\rm old} + 1.53$, 
we expect the relative luminosity values to be accurate to within $\sim$\,0.2~mag.

For absolute magnitudes, galaxy distances are needed, which we base on analyses of
surface brightness fluctuations (SBF; \citealt{Tonry01,Blakeslee09}).
The distances in our sample range from 9\,Mpc (NGC~3115) to 27\,Mpc (NGC 720, NGC~1400 and NGC~1407),
with the median being 17\,Mpc.
The luminosity range is from $M_K = -22.3$ ($L_K = 2 \times10^{10} L_{K,\odot}$; NGC~4474)
to $M_K = - 25.5$ ($L_K = 3 \times10^{11} L_{K,\odot}$; NGC~1407), with
a median of $M_K = -24.0$ ($L_K = 8 \times10^{10} L_{K,\odot}$; NGC~1023).

Further details of the magnitudes and distances are discussed in the Appendix, along with
other parameters presented in Table~\ref{info}, 
including the central velocity dispersion, the effective radius, the morphology,
various isophote parameters, the recession velocity, and environmental parameters.
For some of these parameters (dispersion, etc.), there is an arbitrary choice of spatial scale.
We generally select whatever scale is homogeneously available from the literature (including \atlas3d\
in particular) while being as close to 1\,\Reff\ as possible.

\begin{figure}
\includegraphics[width=\columnwidth]{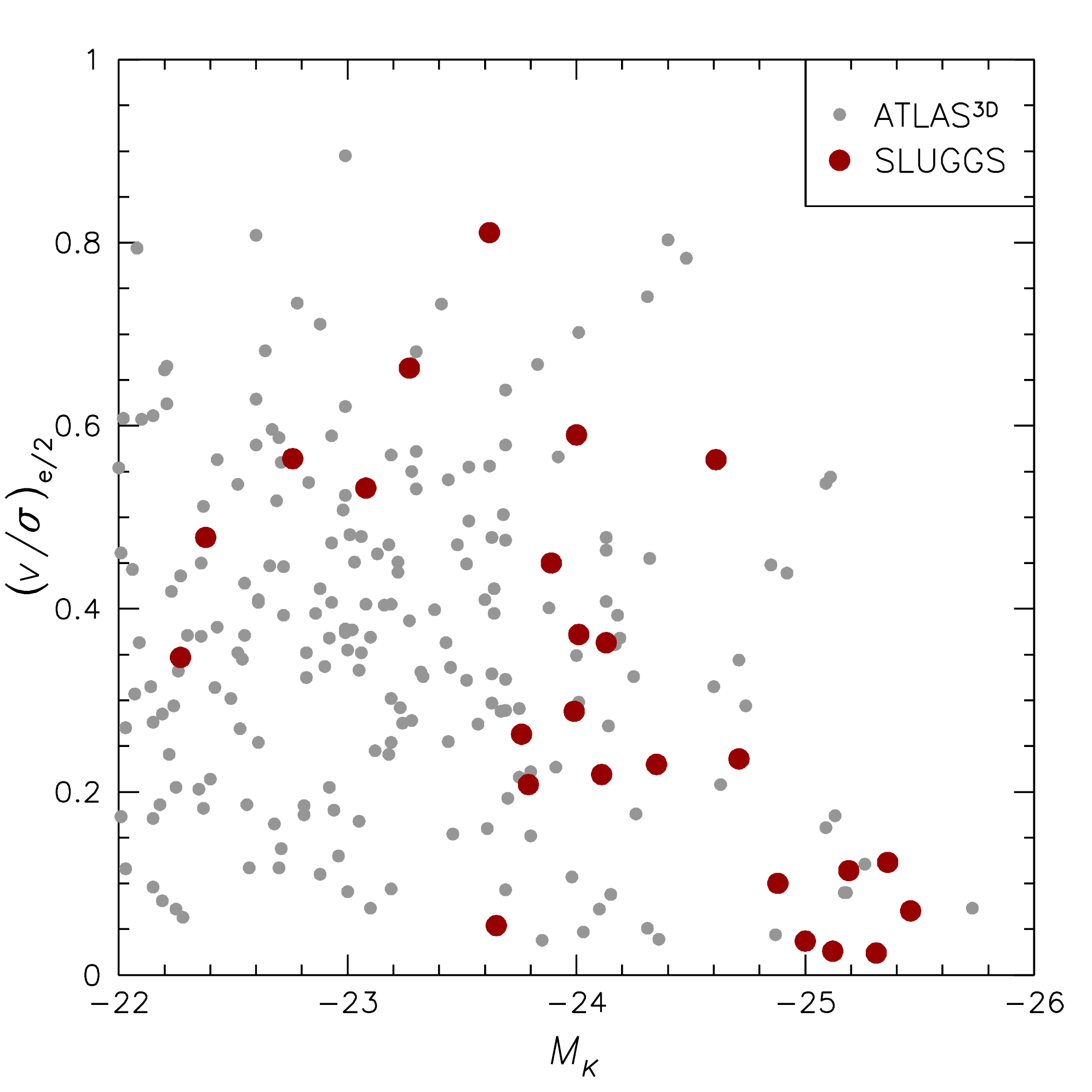}
\caption{Rotation--luminosity space for early-type galaxies in the nearby universe.
Large red points are from the SLUGGS survey; smaller gray points are from \atlas3d.
$M_K \simeq -24.0$ corresponds to the characteristic luminosity $L^*_K$.
Roughly half of the sample are ordinary $L^*_K$ galaxies with a range of apparent rotation properties,
a quarter are very luminous slow rotators, and a quarter are low-luminosity fast rotators.
 }\label{fig:params0}
\end{figure}

As a concise summary of the galaxy sample selection,
Figure~\ref{fig:params0} shows the parameter space of luminosity and rotation,
marking both the SLUGGS galaxies, and the full \atlas3d\ sample for reference.

Here the classic distinction between slow and fast rotators is evident, with almost all
galaxies above and below $\sim 2 \, L^*$ luminosity having $V/\sigma$ less than and greater
than 0.15, respectively.  Our sample includes a large fraction of the luminous slow rotators -- which
have been the classic target of GC spectroscopy for decades -- while now also extending to $L^*$ luminosities and below.
Among the less luminous galaxies, our sample has a bias toward the more rotationally-dominated systems.
This is effectively a selection effect for more edge-on inclinations that we adopted in order to reduce the ambiguity with galaxy deprojection.

The remainder of the galaxy parameters are discussed and diagramed in the Appendix, where it can be seen that
SLUGGS samples the parameter space very well, with only minor quirks such as the inclination bias and
an over-representation of ellipticals at the expense of lenticulars.
We also have a mild selection effect on galaxy recession velocity, requiring 
$V_{\rm sys} \gsim$\,600\kms\ in order to minimize potential confusion between GCs and Galactic stars.
This constraint is mainly relevant to the Virgo cluster, where the intrinsic velocity dispersion leads to some cases of
very low $V_{\rm sys}$, including several well-known giant ellipticals which are therefore not included in SLUGGS:
NGC~4406 (M86), NGC~4472 (M49), NGC~4552 (M89), and NGC~4621 (M59).

The sample size of 25 is approximately motivated as follows.
There are a minimum of three important parameter choices -- bright and faint, cluster and non-cluster, fast and slow rotation, which define eight basic galaxy bins ($2^3$), 
and we have, somewhat arbitrarily, decided that we need at least three galaxies per bin.  
Figure~\ref{fig:thumbs} provides images of all the galaxies on a physically consistent spatial scale. 
Note that there are three additional galaxies whose properties appear at the bottom of Table~\ref{info}. 
They are bonus galaxies that have been observed and analyzed in the same manner as those in the main survey, 
but they were not included in the original sample and have less extensive datasets.  We list them here for completeness.
Other, smaller galaxies are also located within the survey footprints and may be studied serendipitously (e.g., NGC~4283, NGC~4468, NGC~5845).

Out of the 25 SLUGGS galaxies, 19 have photometric data available from SDSS, and 21 are part of the \atlas3d\ survey, including a subset
of 15 galaxies from the original SAURON survey \citep{deZeeuw02} -- which was by design in order to benefit from synergies of combining the datasets
(there are similarly 18 galaxies overlapping with planetary nebula surveys; e.g., \citealt{Coccato09,Teodorescu11,Cortesi13a}).
At the time of writing, DEIMOS spectroscopy has been completed for all but three of the SLUGGS galaxies.

\begin{figure}
\includegraphics[width=\columnwidth]{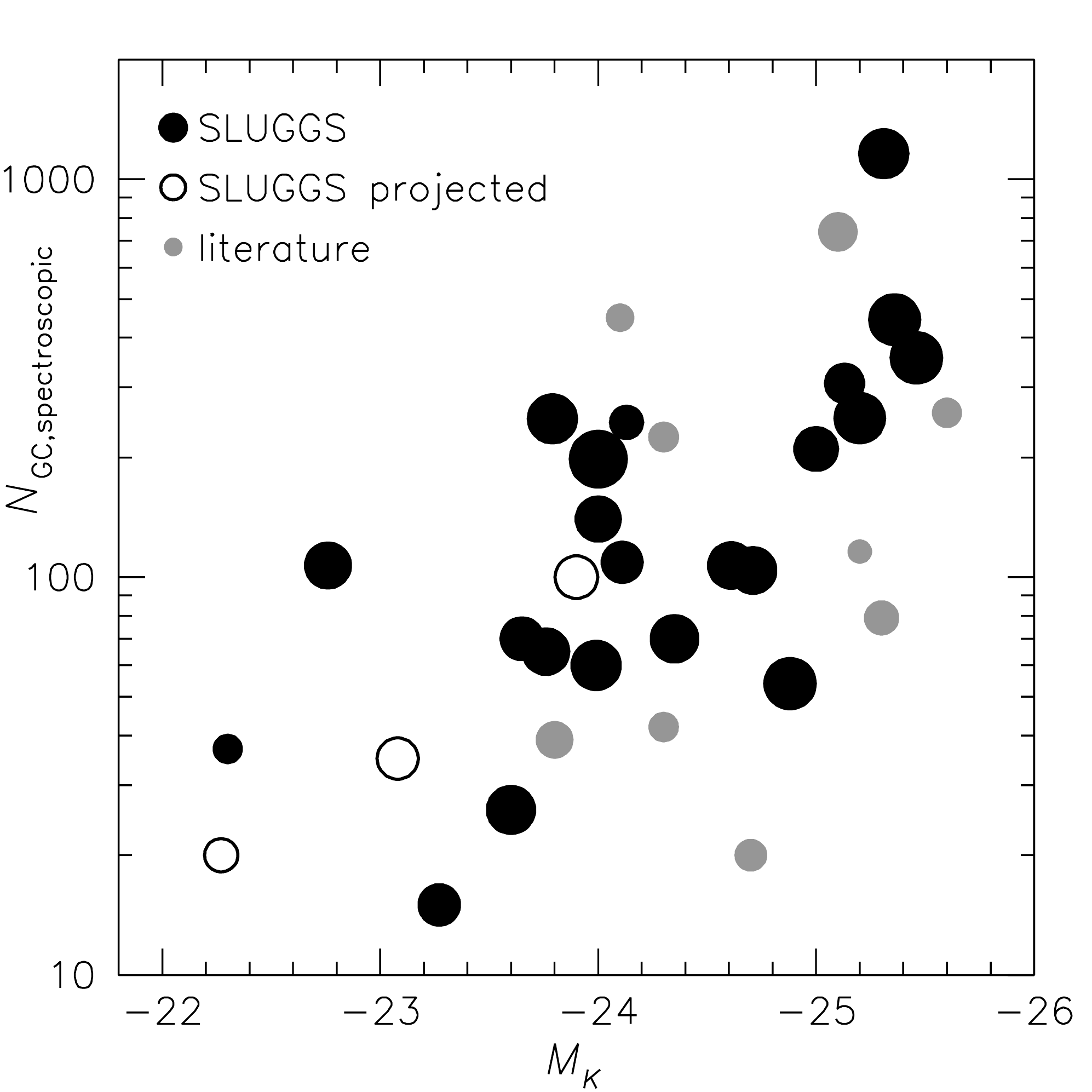}
\caption{Quality and scope of the SLUGGS spectroscopic sample. 
The plot shows the number of spectroscopically confirmed GCs vs.\ absolute magnitude in the $K$-band. 
The symbol sizes are scaled by the velocity precision (bigger is better: 
the scaling uses a logarithmic function of galaxy velocity dispersion divided by GC velocity measurement uncertainty).
Filled black circles denote SLUGGS galaxies already observed and analyzed
(including in some cases supplementary data from other instruments besides DEIMOS),
 while open circles show our projected returns for the remaining survey sample.
 Gray circles show all ETGs from the literature with more than 20 velocity measurements (see \citealt{Pota13a} for references; 
note that the one well-studied $L^*$ case is the nearby peculiar galaxy NGC~5128).
The survey will obtain high-precision velocities of ordinary GCs in large numbers, over a wide range of galaxy luminosities.
 }\label{fig:nspec}
\end{figure}

As a summary of one of the main achievements planned for the survey,
Figure~\ref{fig:nspec} shows the numbers of GCs with spectroscopy, as a function of host galaxy $M_K$ --
including a slight extrapolation for data anticipated as well as already analyzed.
One sees that the $L^*$ galaxies should each provide $\sim$100 radial velocities,
ranging to many hundreds for  the largest galaxies.
The symbol sizes also re-emphasize the importance of velocity precision -- for allowing dynamical studies of low-mass
galaxies, for measuring low rotation amplitudes, and for detecting kinematically cold disks and substructures.

An important feature of SLUGGS is that most of the targeted objects (particularly for kinematics) are ``ordinary'' GCs, 
as opposed to the typical past selection of the brightest tip of the GC population. Many of these luminous objects are ultracompact dwarfs that
have different kinematics and stellar populations than the general GC population (Section~\ref{sec:ucds}).
The unrivaled efficiency of Keck/DEIMOS is a game-changer in this arena: for example,
Gemini/GMOS requires 2.5 times longer aperture-adjusted exposure times for a $\sim$\,1.5\,mag shallower limit,
over a $\sim$\,2.5 times smaller field of view \citep{Bridges06}; 
see also discussion in Section~\ref{sec:gcspec} about velocity reliability issues with other instruments.
This order-of-magnitude improvement
permits for the first time large spectroscopic datasets of ordinary GCs to be obtained around a large sample of galaxies.

\section{Data Acquisition, Reduction, and Analysis}\label{sec:obs}

\begin{figure*}
\centering{
\includegraphics[width=0.8\textwidth]{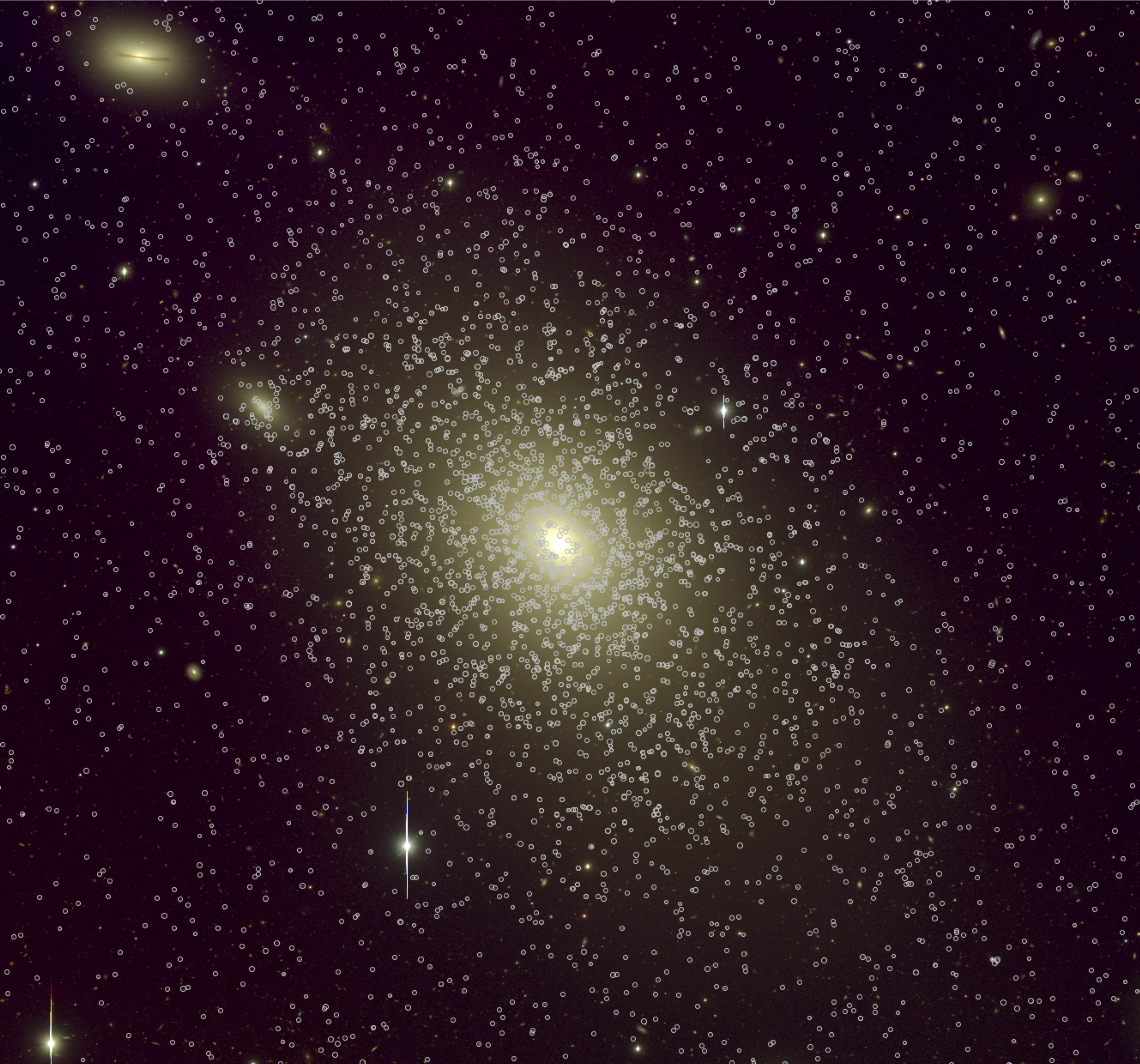}
}
\caption{A three color ($gri$) Suprime-Cam  image of NGC 4365, with its globular cluster (GC) candidates marked by small circles.
This image is a zoom-in at $\sim$\,18$^\prime \times 17^\prime$ ($\sim$\,120\,$\times$\,110\,kpc)
of the original, which is three times the area.
An {\it HST}/Advanced Camera for Surveys image mosaic was also used to select GCs out to $\sim 4^\prime$ from the galactic center.
\citet{Blom12a} determined that NGC~4365 has 6450$\pm$110 GCs and that its GC system extends beyond 9.5 galaxy effective radii.
 }\label{fig:mosaic}
\end{figure*}

The central instruments used for SLUGGS are Suprime-Cam on the 8.2-m Subaru telescope
\citep{Miyazaki02} for imaging, and DEIMOS on the 10-m Keck~II telescope \citep{Faber03} for spectroscopy. 
Here we discuss the techniques for observation and data reduction, leading to the following six basic data products from our survey:
multi-color galaxy surface photometry, GC photometric catalogs, GC radial velocities, GC CaT line-strength measurements,
galaxy-light kinematic moments, and galaxy-light CaT line-strength measurements. 

\subsubsection{Background}
\subsection{Imaging}
\label{sec:Imaging}

High-quality imaging is an important first step for our survey, required for selecting GC candidates for follow up spectroscopy.
However, it is also critical for deriving robust GC surface density distributions (in 2D, including both radial profiles and ellipticities).
These spatial distributions are necessary for estimating the total numbers of GCs ($N_{\rm GC}$) around each galaxy 
and are vital ingredients in models that use GCs as kinematic tracers.  

The GC systems of massive ETGs typically extend to projected radii greater than 100~kpc, and require 
wide-field imaging covering tens of arcminutes on a side in order to obtain reasonably complete spatial coverage.
Therefore the early photographic surveys of nearby galaxies (e.g., \citealt{Harris79}) remained the state of the art for decades until
modern CCD cameras reached the requisite field sizes (e.g., \citealt{Rhode01,Dirsch03,Peng04b}).

Some of this work emphasized the use of three-band (two-color) photometry in order to reduce the contamination of the GC samples by foreground stars and
background galaxies -- a problem which can otherwise become severe at large radii and for the less luminous galaxies.
This issue can be addressed even further by subarcsecond image quality, which resolves out many of the background contaminants.
Deep exposures are also important to reach beyond the peak of the GC luminosity function and thereby to allow for proper GC number counts.

This critical combination of imaging attributes --
wide-field, deep, good seeing, multi-color -- has never before been carried out in a homogeneous survey of galaxies,
but is now an integral part of both NGVS and SLUGGS. The main imaging source for  SLUGGS is Suprime-Cam:
the best instrument  in the world for producing spatially-complete GC system catalogs, owing to 
the telescope's 8\,m aperture, the large areal coverage ($34\arcmin\times27\arcmin$), 
and the typically excellent seeing on Mauna Kea. Most of the data are taken explicitly for our survey,
although some of the images are found in the SMOKA archive \citep{Baba02}.  For a few of the galaxies, we make use
of archival data from CFHT/MegaCam \citep{Boulade03}. In almost all cases these wide-field data are supplemented with
{\it HST} imaging of the galaxy centers (see Section~\ref{sec:gcphot}).

\subsubsection{Imaging Acquisition and Reduction}

Imaging is carried out in three filters for each galaxy: $g$, $r$ and $i$
(except for a few galaxies with adequate archival data available).
Our nominal target for each filter is S/N~$\sim20$ at one magnitude fainter than the turnover of the GC luminosity function (occurring at $M_i\sim-8$). 
The total exposure times depend on the band, galaxy distance, and observing conditions, ranging from $\sim$\,300\,s for $r$-band in the nearest
galaxies in the sample, up to $\sim$\,4000\,s for $g$-band in the most distant. We use a five-point dither pattern to fill in the chip gaps.
We also take very brief, $\sim$~10~sec exposures in order to obtain unsaturated images of the galaxy centers and of any bright
ultracompact dwarfs.

Our reduction of Suprime-Cam imaging is largely based on the SDFRED package \citep{Yagi02, Ouchi04}, but incorporates custom modifications to improve the sky subtraction and alignment of images. High-quality astrometric solutions across the Suprime-Cam field of view are essential for spectroscopic follow-up. In order of preference, astrometric calibration is linked to the Sloan Digital Sky Survey DR7 \citep{Abazajian09}, USNO-B2 \citep{Monet03}, or 2MASS \citep{Skrutskie06}, with typical rms values of $\sim0.15\arcsec$.
An example of a reduced Suprime-Cam image is provided in Figure~\ref{fig:mosaic}.

\subsubsection{Globular Cluster Photometry and Selection}\label{sec:gcphot}

Object detection and photometry are performed using the {\tt DAOPHOT} suite \citep{Stetson92}. All photometry is aperture photometry, using optimal extraction radii and aperture corrections calculated for each set of images. Photometric calibration is achieved using standards taken during the run, or by reference to point sources in SDSS DR7. Corrections for foreground dust extinction are applied using standard maps \citep{Schlegel98}. 

\begin{figure}
\includegraphics[width=\columnwidth]{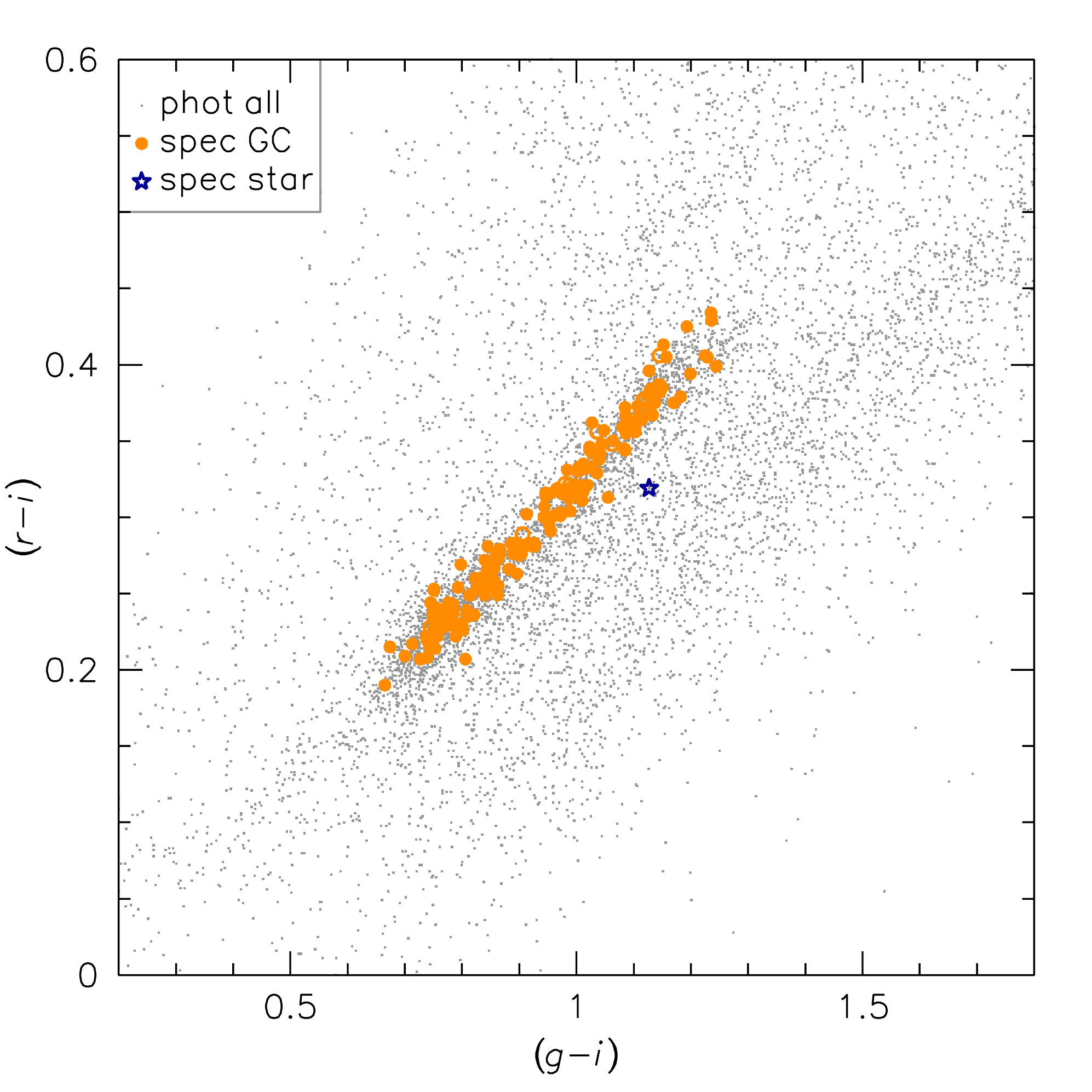}
\caption{Color--color diagram ($r-i$ vs.\ $g-i$) for GCs in NCC~1407.
All objects in the Suprime-Cam field are given as gray points. Objects with spectroscopic identification \citep{Romanowsky09} 
are shown with larger colored symbols: GCs (orange circles) and a star (blue star symbol).
With deep, high-quality imaging and three-band selection, we achieved $< 1\%$ contamination levels in our spectroscopic sample.
Some other galaxies in our sample have sparser GC systems and shallower imaging data, so the results are not as good.}\label{phot_colorcolor_fig}
\end{figure}

Significantly extended objects are removed as probable background galaxies (slightly resolved objects could be ultracompact dwarfs and are not excluded). The GC sequence is defined in $g-i$ versus $r-i$ color--color space (see example in Figure \ref{phot_colorcolor_fig}, and cf.\ \citealt{Sinnott10}). 
This sequence is fairly well-separated from the foreground-star sequence, except at the bluest end (where near-infrared imaging would help
in principle; \citealt{Munoz14}). We thereby obtain a sample of GC candidates that is relatively contaminant-free,
as is important both for spectroscopic target selection, and for general photometric studies of the GC systems.
The bright magnitude GC selection limit is set by the expected range
of ultracompact dwarfs \citep[up to $M_i\sim-13$;][]{Brodie11}; the faint limit varies according to the actual depth of the imaging.

Detection, secure identification, and photometric analysis of GCs from ground-based imaging become challenging near the centers of bright galaxies. 
Fortunately, almost all of the SLUGGS galaxies have archival imaging available from {\it HST}, allowing
for improved GC detection and photometry closer to the galactic centers.
The {\it HST} analysis is similar to that for ground-based imaging, although only one color is typically available \citep[see][]{Spitler06,Strader06}. 
The principal difference is that the vast majority of the GCs in our survey are resolvable by {\it HST}, which allows for another step in
contaminant reduction (eliminating the foreground stars in the central regions of the galaxies).
We also use information from \emph{HST}  to fine-tune the parameters used in the Suprime-Cam GC selection.

Once all of the data have been processed, full photometric catalogs and imaging data for the SLUGGS galaxies will be made publicly available on the project website\footnote{\tt http://sluggs.ucolick.org}.

\subsubsection{Galaxy Surface Photometry}\label{sec:surfphot}

Our wide-field imaging is also suitable for studying the host galaxy,
although this is currently the least advanced aspect of our survey.
Suprime-Cam is not optimally suited for mapping out the extended low-surface brightness regions of galaxies, owing to limitations
with flat-fielding and scattered light that are typical for most telescopes, and require special instrument configurations or
heroic processing efforts to circumvent (e.g., \citealt{Tal09,Janowiecki10,Paudel13}). However, our survey imaging
still reaches fainter surface brightness levels than SDSS, which is important for studying the interesting outer regions of the galaxies.

\begin{figure*}
\centering{
\includegraphics[width=7in]{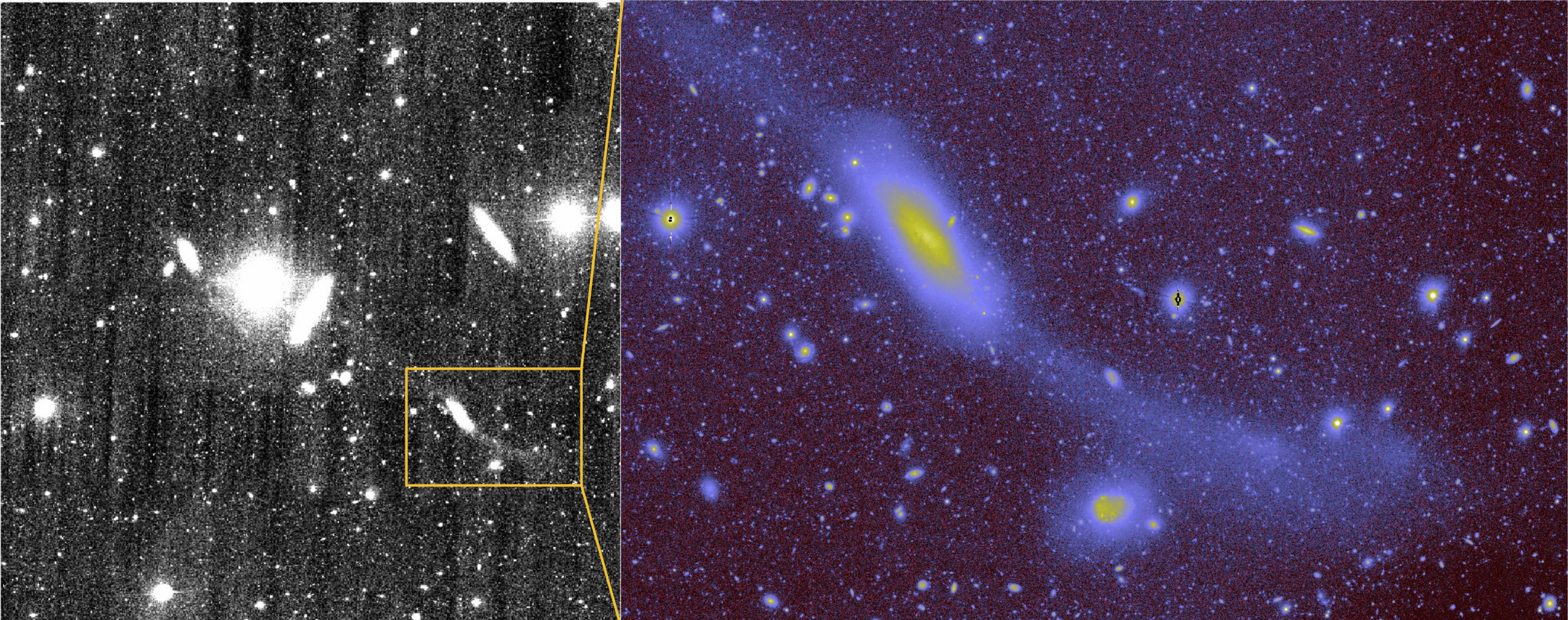} 
\caption{Illustration of the power of Suprime-Cam for studying diffuse  stellar substructures.
The left panel shows an SDSS image, courtesy of Tomer Tal, composed of stacked $g,r,i$ bands and centered on the low-luminosity S0 galaxy, NGC~4111. North is up, East is left (the bright object to its North East is a foreground star).
There is a faint indication of a previously unreported stellar stream extending out 
to the South West through a smaller spiral galaxy.
The area in the box is shown in the right panel with $g$ filter imaging from Suprime-Cam
(with false color used to enhance the image contrast).
Compared to SDSS, the Suprime-Cam imaging has greater depth and resolution, making it possible to analyze the luminosity and detailed structure of the stellar stream, and to do spectroscopic follow up of discrete tracers within it.}
\label{fig:n4111}
}
\end{figure*}

Using the standard IRAF task ELLIPSE, we measure the isophotes in annuli, 
thus building up a 2D description of the galaxy light (see \citealt{Blom12a}). The resulting data include
the radial profiles of surface brightness, ellipticity, position angle, and higher order Fourier components (such as boxyness and diskyness). 
These profiles are used throughout the survey in various contexts, such as in dynamical models, 
derivations of galaxy color profiles, and in comparisons of GC system properties (spatial and kinematical) to the diffuse galaxy light.

The extended stellar light can also be inspected for substructures produced by galaxy interactions and mergers.
Although our sample was designed to avoid galaxies where such features are dominant, we have serendipitously discovered several
halo substructures in our Suprime-Cam data. 

Figure~\ref{fig:n4111} shows one example, presented here for the first time.
An S0 galaxy, NGC~4111, is seen to be linked by a faint tidal stream to a smaller spiral galaxy, UGC 7094 (= PGC 38375). 
The UGC~7094 stream has a maximum surface brightness of $\mu_g \sim 26.4$\,mag\,arcsec$^{-2}$,
spans an apparent length of $\sim$\,75\,kpc, has a slightly redder color than its parent galaxy,
$(g-i)_0 \sim 0.93$ vs.\ $\sim 0.81$, 
and a similar luminosity ($L_g \sim 2\times10^8 \, L_{g,\odot}$ vs.\ $\sim 3\times10^8 \, L_{g,\odot}$).

The stream thus appears to represent an ongoing minor merger, with a stellar mass ratio of $\sim$\,1:20
(based on $i$-band luminosities).
Another S0 galaxy toward the North East, NGC~4117, shows distorted outer features (not shown in the Figure),
suggesting that we may be witnessing multiple interaction events happening within a small group of galaxies as it falls into the Virgo cluster.

The Suprime-Cam imaging also allows us to do far more than characterize substructures photometrically; we can
identify embedded GCs for spectroscopic follow-up with DEIMOS, allowing the chemodynamics of the features to be explored.
So far, we have used this approach for stellar streams around two galaxies in the survey, M87 and NGC~4365 \citep{Romanowsky12,Blom14}. These data allowed us to identify, for the first time, NGC 4342 as the origin of a stream in the NGC~4365 system.

The number of discrete spectroscopic tracers in substructure can be increased beyond GCs by acquiring supplementary narrow-band imaging with Suprime-Cam to identify PNe, which we then follow up with DEIMOS (Foster et al.~2014, in preparation).

\subsection{Spectroscopy}

\subsubsection{DEIMOS Mask Design and Observations}

Our default strategy for spectroscopy is to obtain four $15\arcmin \times 6\arcmin$ Keck/DEIMOS slitmasks for each galaxy in a ``hub and spoke" layout (see figure~3
of \citealt{Arnold14}). This arrangement provides good radial and azimuthal coverage while allowing for mask overlap in the densely populated central regions of the galaxy. In galaxies with especially sparse or populous GC systems, this default of four masks is sometimes adjusted.

Candidates are selected to maximize the number of high-S/N GC spectra per mask. Objects down to $i \sim 23$ are targeted (corresponding to $M_i \sim -8$ for a reference galaxy at the distance of the Virgo Cluster). 
Any unused mask ``real estate" is then devoted to ``filler'' slits (generally centered on random objects) 
used to measure both sky background and galaxy light (see Section~\ref{sec:Stellar spectroscopy} 
below). Typically 50--150 candidates are observed per mask, depending on the mask position and GC system richness. Each mask is observed for 1.5--2.5 hours in individual exposures of 30 min. The total observing time is adjusted depending on the prevailing seeing and transparency conditions.

The observations are made with DEIMOS configured with a 1200 line~mm$^{-1}$ grating centered on $7800$ \AA. The width of the mask slits is 1\arcsec, which implies an average FHWM resolution of 1.5 \AA. This setting provides wavelength coverage of the region $\sim$\,6500--9100\,\AA. 

\subsubsection{Globular Cluster Spectroscopy}\label{sec:gcspec}

The DEIMOS GC spectra are reduced using the {\tt spec2d} pipeline \citep{Cooper12,Newman13a}, which provides flat fielding, wavelength calibration, sky subtraction, and optimized spectral extraction. 

Radial velocities are determined through cross-correlation with a library of stellar templates observed with the same instrument and set-up,
using the region of the CaT at 8498\,\AA, 8542\,\AA, and 8662\,\AA. 
Visual identification of at least two clean lines (usually CaT lines, but sometimes including H$\alpha$),
by at least two independent observers, is required to confirm the velocity. 
Uncertainties are determined using the width of the cross-correlation peak and the variance among stellar templates, with a minimum uncertainty of 5 \kms\ (enforced) and a median value of $\sim$12 \kms\ (determined from the 21 galaxies analyzed so far). Our uncertainties have been verified by repeated measurements over a large range in S/N, which is part of a general framework of redundancy in our program, designed to prioritize quality control
along with sheer quantity of GC velocities.

We note that for a few galaxies, we supplement the spectroscopic coverage with data from other telescopes or instruments
(e.g., Keck/LRIS; MMT/Hectospec; Magellan/IMACS). These additions are discussed in the relevant papers (e.g., \citealt{Arnold11,Strader11}).
Having occasional spectroscopic targets in common between DEIMOS and other instruments, and between multiple DEIMOS observations,
allows us to monitor our results for reproducibility.  
As discussed previously \citep{Romanowsky09,Strader11,Pota13a}, such comparisons demonstrate extremely good consistency for the DEIMOS velocities.
However, a few objects in common with the literature show catastrophically large discrepancies, at up to 10\,$\sigma$ difference, which may be
attributable to marginal S/N in the older measurements.  As discussed in \citet{Strader11}, even a few such outliers can change the
kinematic inferences enormously, which further motivates the strict quality control that we have adopted in our survey.

In typical seeing of 0.8\arcsec, radial velocities are measurable with DEIMOS for a good fraction of the faintest ($i \sim 23$) GC candidates in each mask. The GC return rate per mask depends upon a number of factors, such as the seeing, the intrinsic richness of the GC system, the galactocentric distance of the mask, and the photometric GC selection criteria. 
In the best cases, where the galaxy has a populous GC system and imaging data are available from both {\it HST} and Subaru,
a single mask can return up to 60--70 GCs.
The radial extent of the GC velocities, based on the 12 galaxies analyzed to date \citep{Pota13a}, is typically 8\,\Reff, and in some cases up to 15\,\Reff.

Background galaxies can often be identified by the presence of emission lines in their spectra.
By contrast, foreground G and K stars have spectra that are very similar to those of GCs. In some cases their velocities are sufficient to identify these objects as stars. In rare cases the systemic velocity of the targeted galaxy is low enough that there is potential overlap between the velocity distributions of Milky Way stars and target galaxy GCs.
No universal criteria are used for GC discrimination in these galaxies; uncertain objects have been flagged in the appropriate catalogs and additional discussion is present in the corresponding papers when necessary.

In addition to kinematics,
DEIMOS spectra are used to obtain GC metallicities based on the CaT index which is a measure of the strength of the Calcium~{\small II} triplet 
absorption lines at 8498, 8542, and 8662\,\AA, 
relative to the local continuum; the precise definition of the index may be found in \citet{Usher12}.
Unfortunately, the CaT is in a region of the spectrum containing several strong sky emission lines that make line index measurements difficult for faint objects, such as distant GCs.
We use a technique described in  \citet[][see also \citealt{Usher12}]{Foster10}  to reduce the effect of these sky lines on the CaT index measurement.
We mask the sky line regions and fit a linear combination of stellar template spectra (the same templates used to measure the radial velocity).
Each fitted spectrum is then continuum-normalized, and the strength of the CaT index is measured on the normalized spectrum.
A Monte Carlo resampling technique is used to derive asymmetric confidence intervals for each index measurement. A number of weaker lines of Fe, Mg, Ti, Na, and other elements are present in the GC spectra, and their strengths can be measured once spectral stacking has increased the S/N (see Section~\ref{sec:bimodal} and Figure~\ref{fig:bimodal};  \citealt{Usher14}).

\subsubsection{Stellar Spectroscopy}
\label{sec:Stellar spectroscopy}

The {\tt spec2d} reduction pipeline generates the subtracted background spectra (loosely referred to as ``sky''). However, because our point-like targets are projected onto the host galaxy, these background spectra are a combination of pure sky and galaxy background light. We use the SKiMS technique 
\citep{Norris08, Proctor09,Foster09,Schuberth12,Richtler14} 
to extract the galaxy spectrum from these background spectra. Briefly, we use a specially defined ``sky index" to identify pure-sky slits at large galactocentric distances (typically 6--7~\Reff), where the contribution of galaxy light is insignificant. 

In earlier SLUGGS work, the pure-sky spectra were co-added and normalized  into a single master sky spectrum that was then scaled for non-local sky subtraction for each slit (\citealt{Proctor09,Foster09,Foster11}). In order to minimize the impact of small variations of the spectral resolution across the spatial axis of the DEIMOS mask, in recent SLUGGS work we have used the penalized Pixel Cross-correlation Fitting ({\sc pPXF}) routine by \citet{Cappellari04} to derive a linear combination of pure-sky spectra that best matches the sky component in each slit (see \citealt{Arnold14}). 

This sky spectrum, individually optimized for each slit, is then subtracted from the corresponding slit, and the remaining signal is the galaxy stellar-light spectrum.
Because these SKiMS spectra are spread in two dimensions around the galaxy, in accordance with the GC targets,
we can thereby effectively use DEIMOS as a sparsely sampled integral-field unit.

\begin{figure}
\centering
\includegraphics[width=\columnwidth]{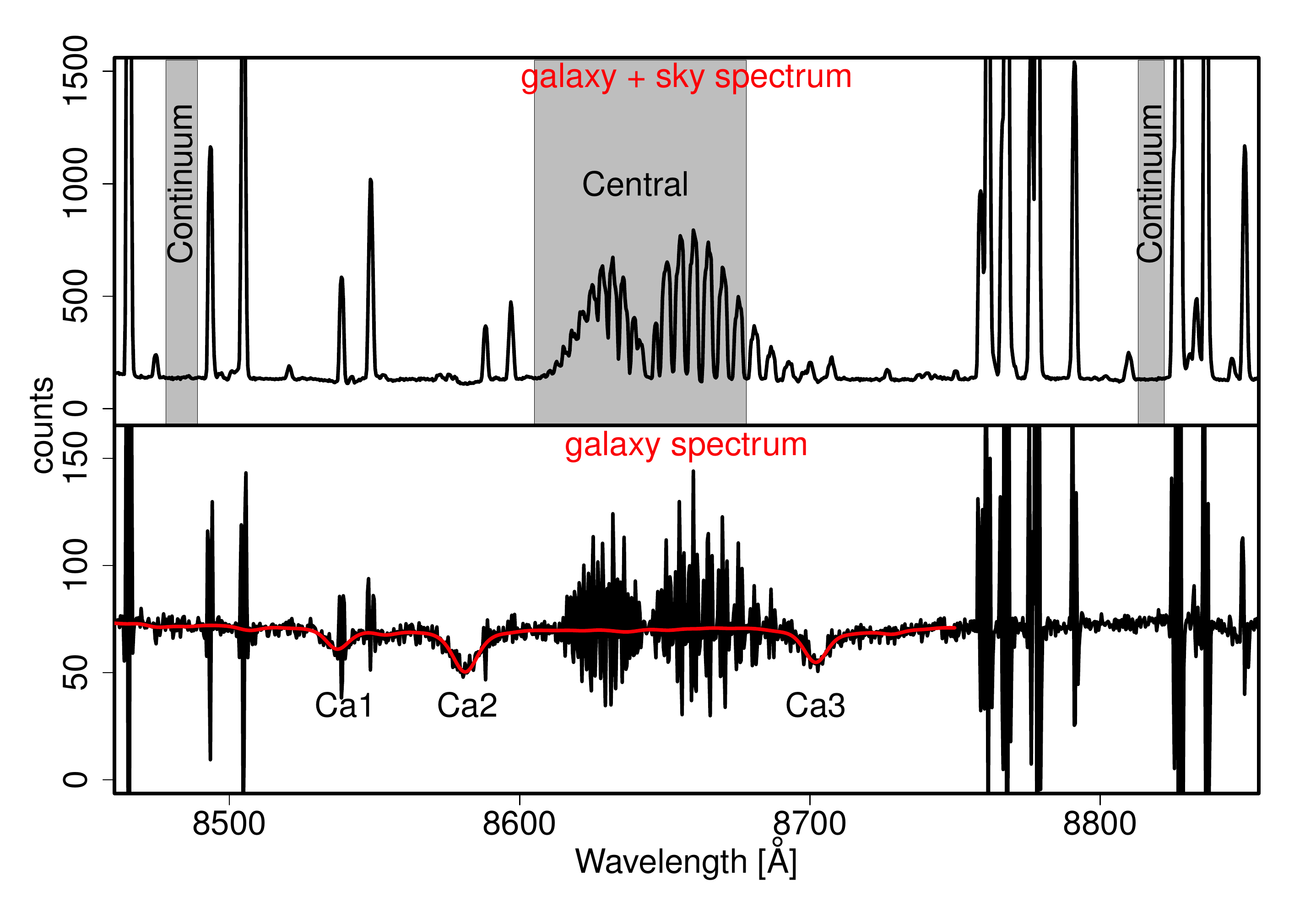}
\caption{Example of a typical DEIMOS spectrum of galaxy light, before (top panel) and after sky subtraction (lower panel), for NGC~4494
at $\sim$\,1\,\Reff, a location where the surface brightness  is $\mu_i=21.4$\,mag\,arcsec$^{-2}$.
The {\sc pPXF} template fit is shown in red. The passbands of the sky index are highlighted in gray in the original spectrum (see text) and the positions of the redshifted CaT features are labeled on the subtracted spectrum.}\label{fig:stellarspec}
\end{figure}

This method yields a final continuum level in the sky index definition region accurate to $< 1$\% of the noise in the skyline residuals \citep[see][]{Foster09}. As an example, Figure \ref{fig:stellarspec} illustrates the sky index definition and sky-subtracted spectrum for the galaxy light of NGC~4494. Using this procedure, we obtain galaxy spectra that typically extend to 2--3\,\Reff, with some reaching 4\,\Reff.

Stellar kinematics are measured using pPXF.
Briefly, the code uses the fully propagated variance array to proportionally ``penalize'' pixels in the spectrum. Template spectra are convolved with Gauss--Hermite polynomials and compared to the observed spectra. The routine identifies the set of kinematic parameters and templates that minimizes residuals between the observed and combined template spectrum. We simultaneously fit the first four velocity moments (i.e., recession velocity, velocity dispersion, $h_3$, and $h_4$). Uncertainties are measured using Monte Carlo methods. Our kinematic measurements are described in detail in \citet{Arnold14}. Figure~\ref{fig:stellarspec} shows an example SKiMS spectrum together with its best fit {\sc pPXF} output template spectrum.

For the highest S/N spectra, we measure the strength of the CaT index for the diffuse stellar light to large radii using the technique described in \citet{Foster09}. The index is the weighted sum of the equivalent width of all three CaT lines. A correction is applied to account for the velocity dispersion broadening. CaT indices are then converted into metallicity as described in Section~\ref{sec:stelmet}. In this manner, we are able to generate metallicity maps out to typically $\sim$\,1.5--2\,\Reff\ , with some reaching 3~\Reff\
(not quite as far as the kinematics measurements owing to the greater S/N requirements;
for more detail, see \citealt{Foster09, Foster11, Pastorello14}).

\section{Modeling}\label{sec:model}

Here we describe various methods for modeling the observational measurements from Section~\ref{sec:obs}
and thereby converting them into products of more direct scientific interest, such as maps of stellar velocities,
mass profiles, etc.

\subsection{Globular Clusters}

\subsubsection{ GC Numbers and Spatial Distributions}\label{sec:gcprofiles}

Accurate GC number counts, including reliable uncertainties, are important for interpretations of their
formation histories (e.g., \citealt{Bekki08,Coenda09}), estimates of stellar halo masses (Brodie \& Strader 2006), and for analyses 
of their correlations with interesting galaxy properties such as supermassive black hole masses and dark matter halos
(e.g., \citealt{Spitler09, Burkert10, Harris11, Snyder11, Rhode12, Harris13, Harris14,Wu13,Forte14,Durrell14}). 
One of the goals of the SLUGGS survey is to provide an updated set of reliable GC numbers
(including the red and blue subpopulations separately), and then to explore what impact these have on the scaling relations. As described
below, some of the literature values of $N_{\rm GC}$ are in error by factors of 3 or more, and we anticipate that our more accurate 
estimates will clarify these fundamental relations. 

Although every care is taken to ensure the GC catalogs from photometric data contain very little contamination (from foreground
stars and background galaxies; see Section~\ref{sec:gcphot}), a certain amount is unavoidable 
and will impact measurements of the true GC number counts.  Fortunately, the contaminants
contribute a fairly constant surface number density over the regions covered by the GC systems.  This surface density is 
typically estimated as a free parameter in fits to the GC surface density profiles, and can be cross-checked via spectroscopic contamination rates \citep{Strader11,Usher13}.

To derive total GC numbers around a galaxy, the surface density profile is integrated with the radius.  
Correction must also be made for the fraction of GCs that were too faint to be included in the catalogs. Since the functional form of the GC system luminosity function is predictable \citep{Jordan07}, the missed GCs can be accounted for by measuring the brighter parts of the luminosity function and extrapolating to faint magnitudes (which can be done securely because the SLUGGS images are designed to reach past the turnover luminosity). Corrections are thus made for completeness and contamination as a function of magnitude.

\begin{figure}
\centering{
\includegraphics[width=\columnwidth]{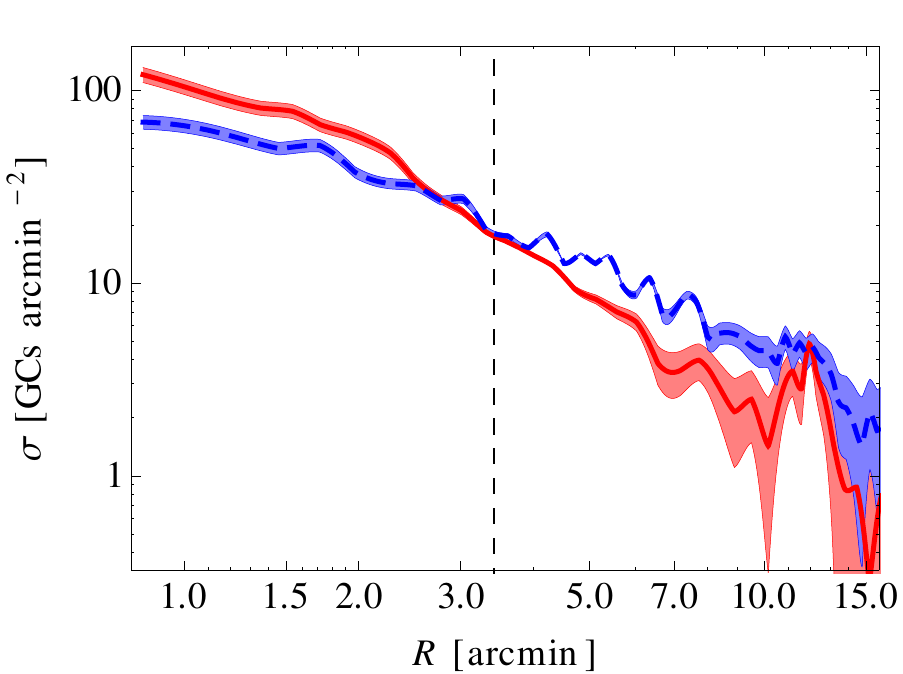}
\vspace{0.1cm}
}
\centering{
\includegraphics[width=\columnwidth]{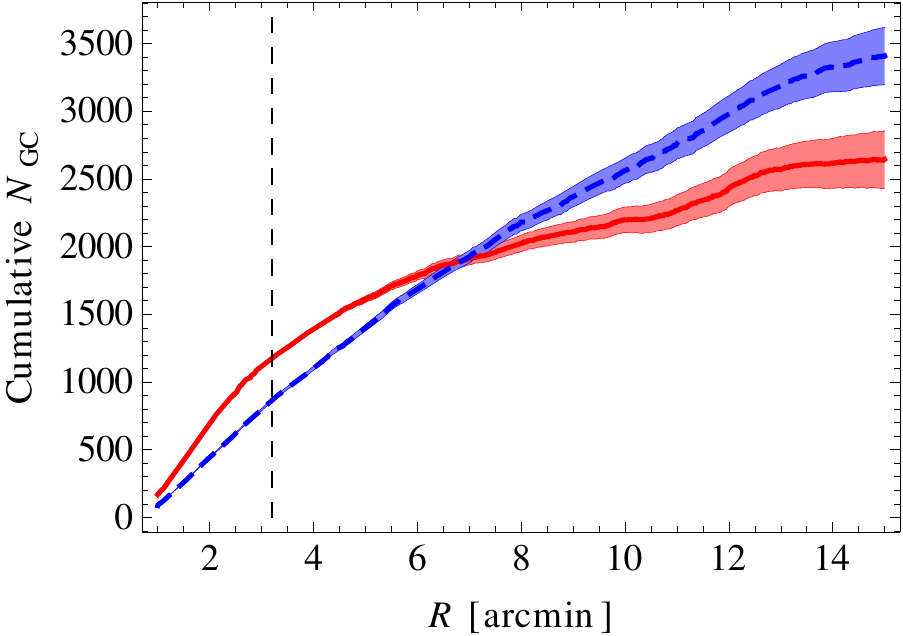}
}
\caption{Radial distribution of GCs around NGC~1407, with the blue and red GC subpopulations shown as 
dashed blue and solid red curves, respectively. 
In both panels, the shaded areas show the 1\,$\sigma$ uncertainty regions, including both Poisson errors and the uncertainties in the
contamination levels (which have been subtracted from the profiles).
Top panel: Interpolated surface density profiles with logarithmic galactocentric radius.
Bottom panel: Cumulative numbers with linear radius.
For simplicity, no attempt has been made to mask out the contributions from the smaller companion galaxy,
NGC~1400, which produces the bump in the red GC profile starting at $\sim$~12 arcmin.  
The vertical dashed line marks the limits of the  (off-center) {\it HST}/ACS field-of-view.
The GCs are selected over the magnitude range of $i=$\,20.6--25.5, with no correction applied for incompleteness over
the luminosity function,
so the numbers here are relative, not absolute. The profiles are terminated when the annuli reach the edges of the field, at about 15 arcmin. 
The cumulative profiles begin to level off at this radius, allowing for reasonably reliable estimates of the total GC numbers by
using the wide-field Suprime-Cam imaging.
The red GC system is more centrally concentrated, whereas the blue GC system is more radially extended, as found for most massive galaxies.
}
\label{fig:gcprofile}
\end{figure}

As an illustration of the importance of reliable surface density distributions for deriving accurate estimates of $N_{\rm GC}$, 
Figure~\ref{fig:gcprofile} shows the surface density and cumulative number profiles of blue and red GCs with radius around NGC~1407, 
based on Suprime-Cam imaging.
The cumulative profiles flatten out at a radius of $\sim$\,14\,arcmin ($\sim$\,110\,kpc), corresponding to the edge of the Suprime-Cam image.
Thus, with very high quality, wide-field imaging, one can just barely get a reliable count of total GC numbers around a high-mass elliptical galaxy at 27 Mpc.

For poorer-quality data, narrower fields, or more nearby massive galaxies, the total GC counts can become precarious.
For NGC~1407,  a single {\it HST}/ACS pointing picks up only $\sim$~10\% of the total GCs
\citep{Forbes06}, and ground-based imagers with $\sim$~10--20~arcmin fields would subtend only half of them.
Naturally such limitations are obvious, and attention has been given in the literature to making reasonable extrapolations 
while accounting for the associated uncertainties, but the results until now have not been encouraging.
As a case in point, the giant elliptical NGC~4365 was analyzed with a central ACS pointing supplemented
by parallel WFPC2 imaging, and estimated to have $3246\pm598$ GCs by \citet{Peng08}.
Our subsequent analysis of Suprime-Cam imaging found a lower-limit of 
$6450\pm110$~GCs \citep{Blom12b}, which is higher than the previous result by at least 5~$\sigma$.

Besides total numbers,
our Subaru data allow accurate surface density distributions  to be derived for the SLUGGS galaxies out to large radii.  We can improve on past procedures for determining important basic parameters for these galaxies, such as \Reff,  by fitting S\'ersic models both to the galaxies and to their GC systems \citep[see e.g.,][for GCs]{Blom12b, Usher13, Kartha14}.

On a related note, the analysis of the radial ``extent'' of GC systems (e.g., \citealt{Rhode07})
is also sensitive to limitations with the imaging data.  For example, analysis of the SLUGGS data for NGC~720
by \citet{Kartha14} revealed that its GC system extent had previously been underestimated by a factor of three.

\subsubsection{GC Kinematics}
\label{sec:GC kinematics}

The kinematics of a GC system can be characterized by the rotation amplitude $v_{\rm rot}$ along a position angle $\theta_0$, and by the rotation-subtracted velocity dispersion $\sigma_{\rm p}$. These parameters quantify the amounts of coherent rotation and random motion of the system, and they can be compared to the same quantities for stellar kinematics.  We study these quantities using a maximum-likelihood approach \citep{Kissler-Patig98b}
that we have extended to allow for flattened galaxies (see details in \citealt{Pota13a}, and discussion in \citealt{Romanowsky09} about
problems with other methods).
The approach is equivalent to a simplified version of kinemetry used for galaxy stellar-light kinematics 
(see Section~\ref{sec:Stellar kinematics}).

\begin{figure}
\includegraphics[width=\columnwidth]{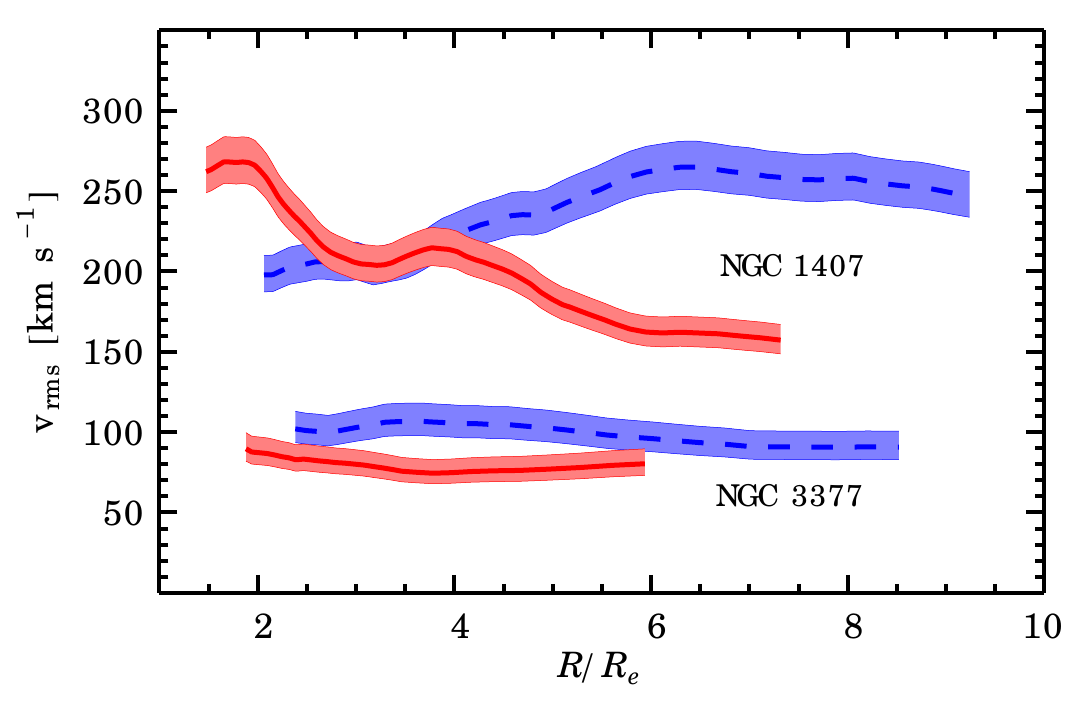}
\caption{GC system rms velocities vs.\ galactocentric radius, normalized by the galaxy effective radius,
for NGC~1407 (top curves) and NGC~3377 (bottom curves). Dashed blue and solid red curves correspond to  
the blue and red GC subpopulations separately. Uncertainties are indicated by shaded regions.
The radii have been circularized to account for galaxy ellipticity (see \citealt{Pota13a}).
The blue and red GCs show kinematical differences that reflect distinct spatial and orbital properties.
}\label{fig:GCSkin}
\end{figure}

One particular kinematic measurement of interest is the projected rms velocity  \vrms\
which, as a metric of specific kinetic energy, can be connected through dynamical modeling to estimates of mass.
Figure~\ref{fig:GCSkin} shows \vrms\ profiles for two example SLUGGS galaxies, 
one high-mass (NGC~1407) and one low-mass (NGC~3377), using SLUGGS data from \citet{Pota13a}.
For each galaxy, the data are separated into profiles for the blue and red GC subpopulations.
The case of NGC~3377 illustrates the generic result that the blue GCs have higher rms velocities than the red GCs,
as expected from the larger spatial extent of the blue GC system density profile (which must be supported by hotter kinematics in
the same gravitational potential).
In some cases like NGC~1407, the radial behaviors of the \vrms\ profiles are also different, which
probably reflects very different orbital types between the blue and red GCs.

Interpreting these kinematic data through dynamics, in terms of orbital properties and mass profiles,
will be discussed further in Section~\ref{sec:mass}.
It should be noted however that the observed \vrms\ profiles are not generally smooth,
which is a complication that should ultimately be addressed by allowing for strong orbital
variations in the models.
In addition, some of the observed fluctuations may be produced by substructures, thereby providing
valuable clues about recent merging events
(cf.\ \citealt{Romanowsky12,Coccato13,Schauer14,Murphy14}).

\begin{figure*}
\centering{
\includegraphics[width=5.8in]{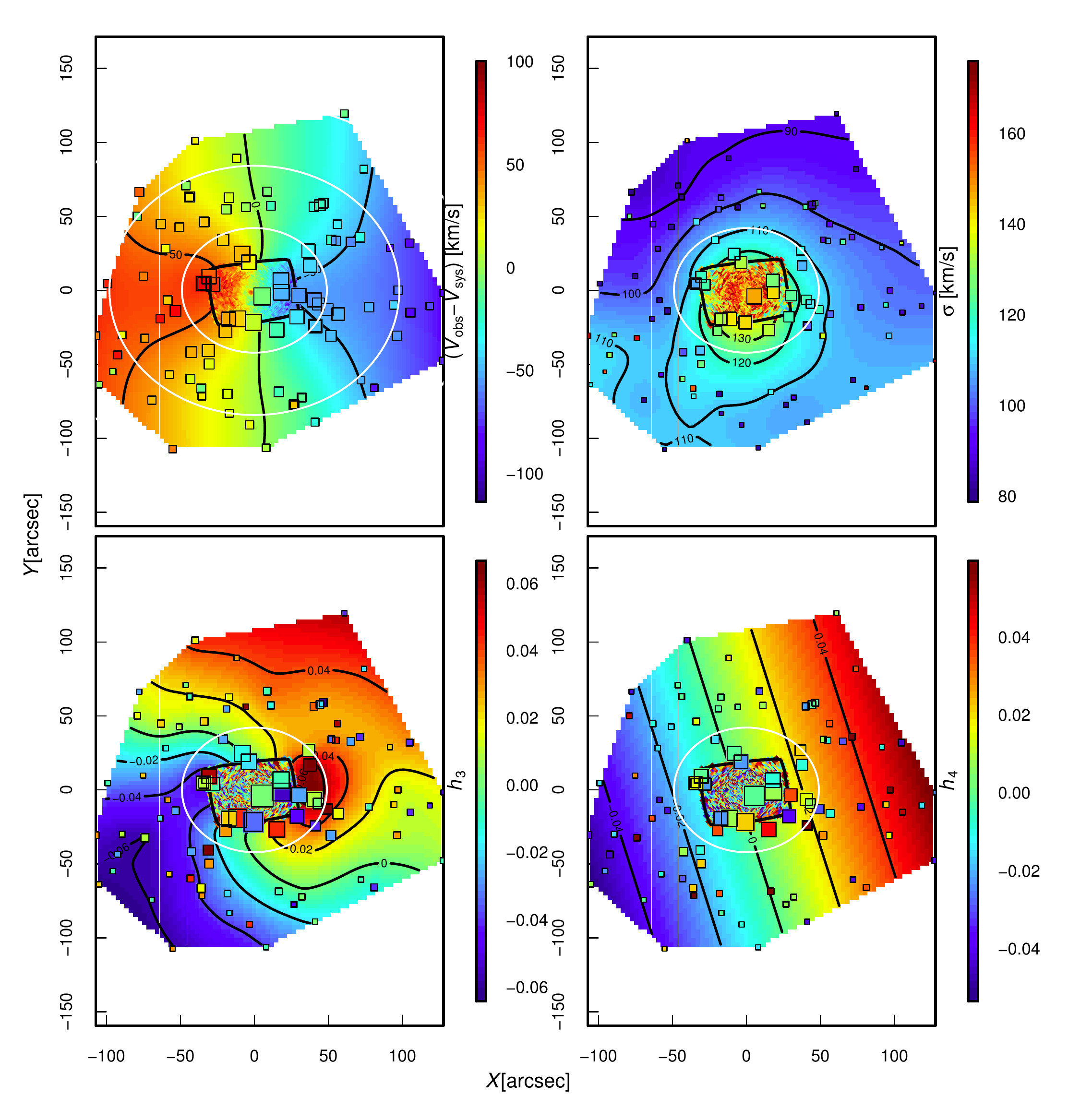}
}
\caption{Example 2D stellar kinematic maps for the elliptical galaxy NGC~4494:
recession velocity (upper left), velocity dispersion (upper right), third-order ($h_3$, lower left) and fourth-order ($h_4$, lower right) Gauss--Hermite
velocity moments. Filled squares show individual SKiMS data points \citep{Foster11}, with sizes proportional to spectral signal-to-noise.
The smooth color mapping and iso-moment contours (black) are derived using kriging to fit the underlying 2D field (see e.g., \citealt{Foster13,Pastorello14};
the color scale is provided to the right of each panel, and the contours are labeled). White ellipses in all panels show the photometric effective radius (\Reff), with the upper left panel also including the 2\,\Reff\ ellipse. 
These maps join smoothly to the SAURON/\atlas3d\ data from the central regions (shown within rectangles), 
but as in Figure~\ref{fig:layer}
show novel features in the outer regions, including a surprising degree of flattening for the outer mean velocity field
(which is not traced by the very round isophotes).
}\label{fig:krig4494}
\end{figure*}

\subsubsection{GC Metallicities}
\label{sec:GC metallicities}
The metallicities of the GCs are derived using the CaT index (Section~\ref{sec:gcspec}), whose strength
has previously been shown to be sensitive to changes in mean metallicity \citep{AZ88,Diaz89,Cenarro01,Foster09,Foster10}.
CaT is independent of age for stellar populations older than 3 Gyr \citep{Vazdekis03}. 
Although the index is sensitive to a bottom heavy initial mass function (IMF; \citealt{Vazdekis03}), 
our CaT measurements are insensitive to horizontal branch morphology and to $\alpha$ element enhancement \citep{Brodie12}.

We have compared CaT metallicities with literature values derived using traditional Lick Index analyses, finding excellent agreement between the two techniques (with a reduced $\chi^2$ value of 0.59; \citealt{Usher12}).
We use the single stellar population models of \citet{Vazdekis03} to derive the following relation for converting the CaT index strengths into metallicities:

\begin{equation}
[Z/\text{H}] = (0.461 \pm 0.013) \, \text{CaT} + (-3.750 \pm 0.080) \, .
\label{eqn:ZH}
\end{equation}

Further details of the CaT metallicity measurement techniques are given in \citet{Usher12}.

\subsection{Galaxy Starlight}

\begin{figure*}
\includegraphics[width=\textwidth]{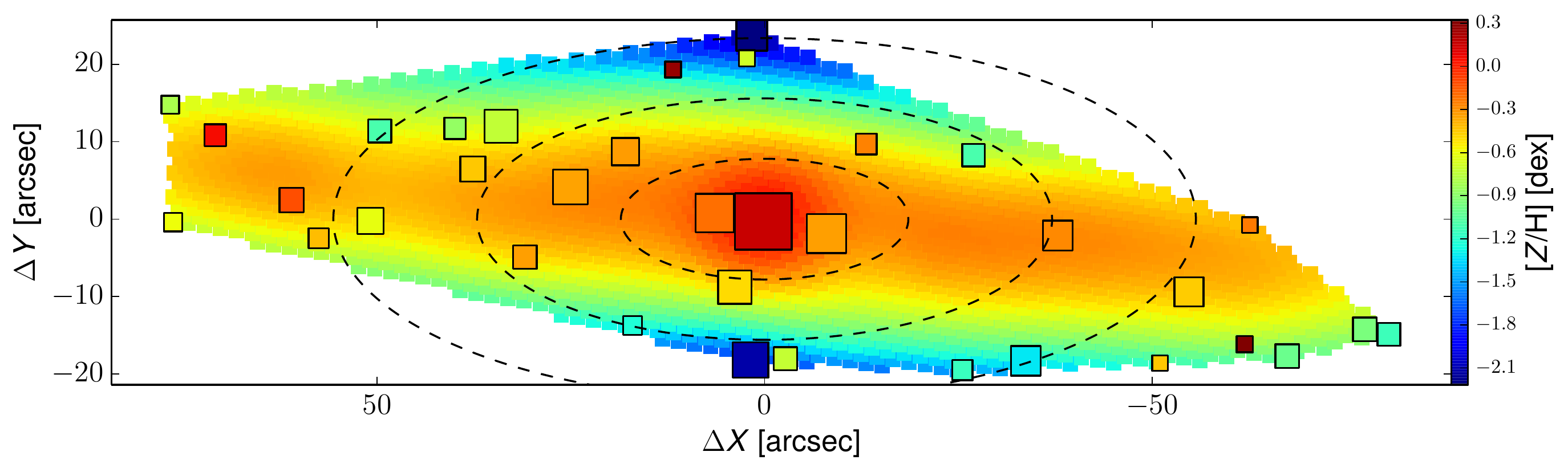}
\caption{NGC~4111 stellar metallicity map (where $X$ is the galaxy major axis, and $Y$ is the minor axis).  The square symbols show the SKiMS metallicity values, color coded according to their metallicity (see scale
bar at right) and with the size inversely proportional to the associated uncertainty.
The pixel values are obtained via kriging interpolation of the SKiMS points and are color coded according to the same metallicity scale. From the center to the outskirts, the dashed lines present the galaxy isophotes with circularized radius $R=1,2,3~R_{\rm{e}}$. The coordinates on the axes are relative to the center of the galaxy.
An extended metal-rich disk is apparent from this map (note that the slight tilt of the disk is an artifact of the incomplete spatial sampling).
A strong negative metallicity gradient emerges along the minor axis.}
\label{fig:metalmap}
\end{figure*}

\subsubsection{Stellar Density Distributions}

As for the GC density distributions (Section~\ref{sec:gcprofiles}), the stellar densities are needed for dynamical models,
after converting the raw density profiles derived from the images (Section~\ref{sec:surfphot}) into parameterized forms
such as the \citet{Sersic68} law.
The photometry can also be analyzed to derive estimates of galaxy luminosity, effective radius, etc.\ -- basic quantities
which as discussed in Section~\ref{sec:samp} and in the Appendix, are reliably and homogeneously available for our sample.
Details of the galaxy photometry and results will be reported in future papers.

\subsubsection{Stellar Kinematics}
\label{sec:Stellar kinematics}

Once the SKiMS-based stellar kinematics measurements are obtained (Section~\ref{sec:Stellar spectroscopy}), the overall kinematic structure of the
galaxy can be visualized through 2D maps.  Several methods have been used for preparing these maps,
including Voronoi binning with semi-variogram-based smoothing \citep{Arnold14}, and
kriging interpolation \citep{Krige51,Matheron63,Cressie88,Foster13}.
An example of the latter is shown by Figure~\ref{fig:krig4494}.

As with the GCs (Section~\ref{sec:GC kinematics}), 
the full 2D kinematical information for the stellar light can be reduced to 1D profiles through modeling.
We adopt  a ``kinemetry'' approach,  originally designed by \citet{Krajnovic06} for IFU data-cubes,
 and extended to sparsely sampled data by \citet{Proctor09} and \citet{Foster11}. 
Kinemetry is analogous to ellipse-based galaxy surface photometry and assumes that the rotation and the dispersion are stratified on ellipses.
These profiles generally include the rotation amplitude, position angle, and ``flattening'' or ellipticity
(which is often, but not always, similar to the photometric ellipticity; cf.\ \citealt{Krajnovic08}).
The kinemetry results for the survey will be presented in Foster et al.\ (in preparation).

\subsubsection{Stellar Metallicities}\label{sec:stelmet}

As discussed in Section~\ref{sec:Stellar spectroscopy},
CaT indices can be measured for the integrated starlight out to radii of typically $\sim$\,1.5--2\,\Reff.
These values are converted to total metallicities [$Z$/H] using Equation~(\ref{eqn:ZH}),
but with an empirical correction that takes into account the observed dependence of CaT-derived metallicities on galaxy mass.
\citet{Pastorello14} suggested that this dependence may be a reflection of the index's sensitivity to the slope of the IMF, 
and derived the correction by comparing our values to metallicities from overlapping SAURON data \citep{Kuntschner10}.

The uncertainties are evaluated using Monte Carlo resampling of the spectra, in a method similar to
the one adopted for GC metallicities (Section~\ref{sec:GC metallicities}).

In order to obtain smooth 2D stellar metallicity maps for the galaxies in our sample, we interpolate 
between our data points with the kriging technique \citep{Pastorello14}.
An example of a 2D stellar metallicity map derived in this way is shown in Figure~\ref{fig:metalmap} for NGC~4111.

From these 2D maps we extract azimuthally averaged radial metallicity profiles, from which 
we measure the inner ($R<1$\,\Reff) and outer ($R>1$\,\Reff) metallicity gradients.  These gradients are powerful tools
 for constraining galaxy formation models and feedback processes (e.g., \citealt{Mihos13}; Hirschmann et al.~2014, in preparation).
 For example, galaxies with more active merger histories should have fairly flat metallicity profiles at large radii \citep{White80},
 while galaxies dominated by in-situ star formation should have steep gradients, with their
chemical enrichment being dependent on the distance from the center of the gravitational potential.

\subsection{Galaxy Dynamics and Masses}\label{sec:mass}

One of the main science drivers for SLUGGS is to use stars and GCs
as dynamical tracers in the outer regions of early-type galaxies and thereby to map out their dark matter halos, 
since H{\sc i} rotation curves are not generally available in these galaxies.
Although this approach, along with the use of PNe, has been pursued for decades,
robust conclusions have remained elusive. This is in part because of
 modeling uncertainties and degeneracies (e.g., \citealt{DeLorenzi09}), but also because the lack of
a large sample of galaxies with homogeneous data has limited the statistical weight of the conclusions.

\begin{figure*}
\includegraphics[width=9.3cm]{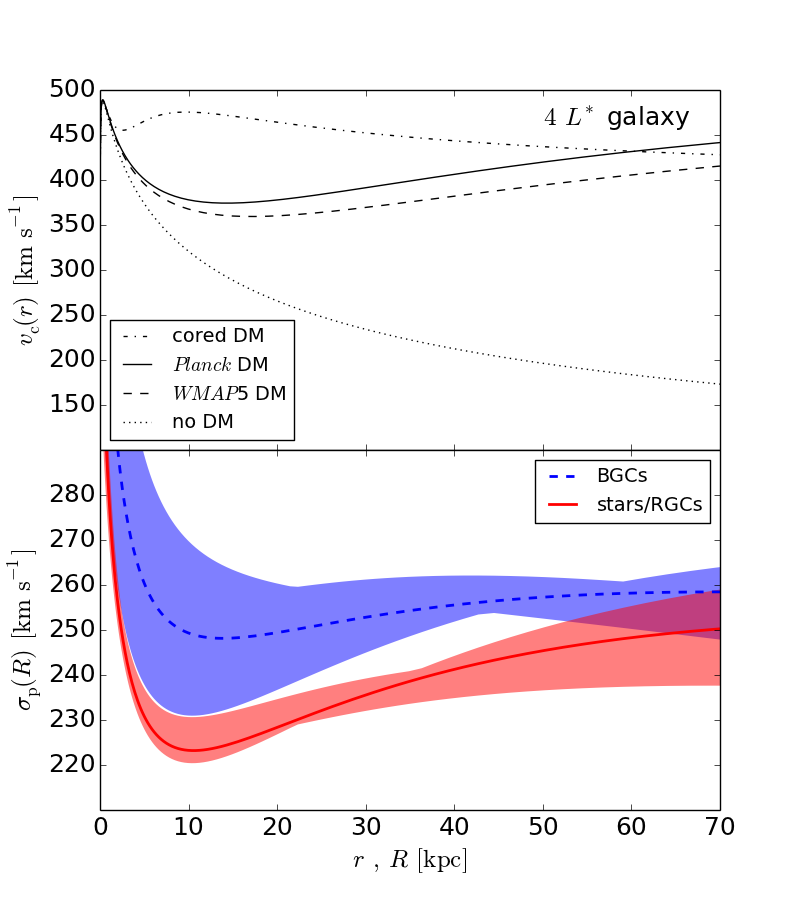}
\includegraphics[width=9.3cm]{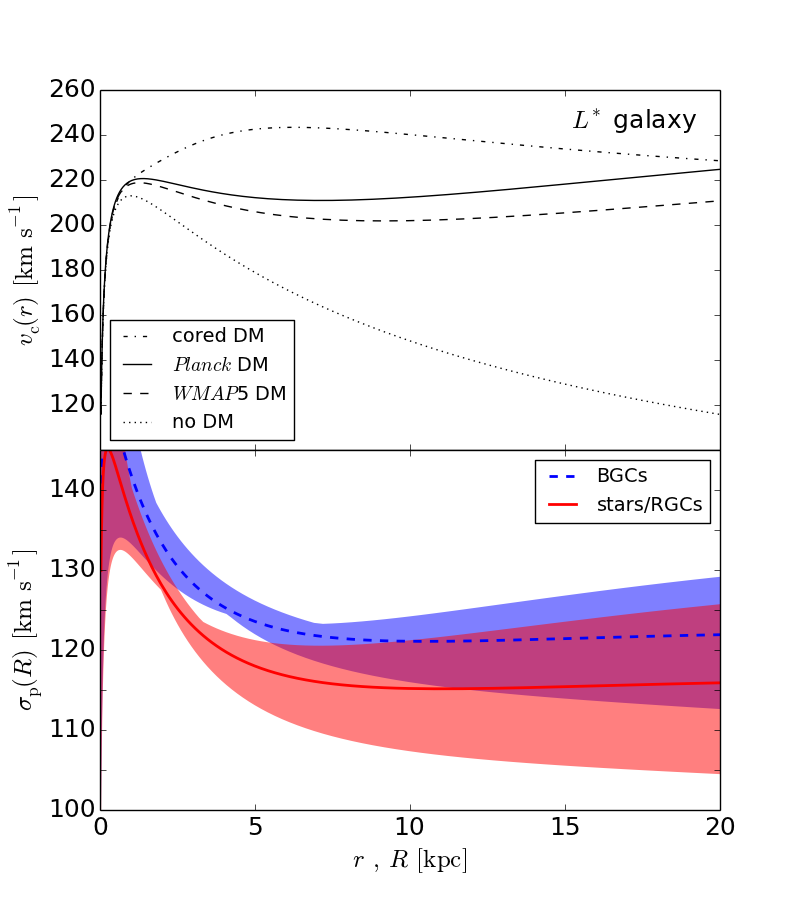}
\caption{Schematic dynamical models for two elliptical galaxies, with $4\,L^*$ (left) and $L^*$ (right) luminosities,
showing radial profiles of circular velocity (top) and projected velocity dispersion (bottom).
The circular velocities cover a range of models including no dark matter, as well as cored dark matter halos, and typical dark matter halos for different
$\Lambda$CDM cosmological parameters (the latter systematic variation is similar to the galaxy-to-galaxy scatter for a fixed cosmology).
These models are very similar inside a few kpc, and diverge at larger radii.
The dispersion profiles are all calculated for the {\it Planck}-based model, with different assumptions on the velocity anisotropy,
including isotropy ($\beta=0$; central curves), tangentially-biased ($\beta=-1$), radially-biased ($\beta=+0.5$), 
and varying with radius from isotropic to radially-biased \citep{Mamon05b}.
 The blue GCs and red GCs (or field stars) are shown separately, with dashed and solid curves, respectively.
For a fixed anisotropy, the blue GCs have a higher velocity dispersion than the red GCs (cf.\ real data in Figure~\ref{fig:GCSkin}).
For each subpopulation, there is a characteristic radius (the ``pinch-point'') where the dispersion profiles cross and the
uncertainty from the anisotropy is minimized (cf.\ \citealt{Agnello14a}; Pota et al., in preparation).
The pinch-points for the more luminous galaxy extend deep into its dark matter halo, providing ample opportunities for model constraints,
while the ordinary galaxy is more challenging to model definitively.
}
\label{fig:toy}
\end{figure*}

Furthermore, the galaxies previously well-studied with GC dynamics were limited to the
high-luminosity ellipticals at the centers of galaxy groups and clusters,
while it is only with the advent of more efficient spectroscopy that the realm of
$\sim L^*$ early-types can be investigated properly (see Figure~\ref{fig:nspec}).
This is an important area of parameter space not only because such galaxies are
more ``ordinary'' and allow for fairer comparisons with $\sim L^*$ spirals such as the Milky Way (e.g., \citealt{TrujilloGomez11}),
but also because of reports based on PNe that the mass profiles of these galaxies are
systematically different from those of the brightest ellipticals, with much less dominant dark matter halos
\citep{Romanowsky03,Napolitano11,Deason12}.

In order to focus the discussion about the limitations and opportunities for dynamical modeling,
we construct a set of toy galaxy models and present the circular velocity and 
projected velocity dispersion profiles in Figure~\ref{fig:toy}.
These models consist of spherical, early-type galaxies, with either $4\,L^*$ or $L^*$ luminosities, 
typical GC systems, and either standard $\Lambda$\,cold dark matter
($\Lambda$CDM) halos \citep{Navarro97b}\footnote{The
$4\,L^*$ galaxy has a stellar mass of $M_\star = 6.6\times10^{11} M_\odot$, and a S\'ersic profile with index $n=8$, and \Reff\,$=15$\,kpc.
The red GC system is assumed to follow the diffuse stellar light, and the blue GC system has $n=4$ and \Reff\,$=50$\,kpc.
The dark matter halo in our default {\it Planck}-based cosmology 
has a scale radius of $r_s = 120$\,kpc and a scale density of $\rho_s = 1.4\times10^6 M_\odot$\,kpc$^{-3}$
(corresponding to virial mass and concentration of $M_{200} = 3.6\times10^{13}M_\odot$ and $c_{200}=4.5$;
\citealt{Dutton14}).
The $L^*$ galaxy has $M_\star = 7.2\times10^{10} M_\odot$,  $n=4$, and \Reff\,$=3.5$\,kpc (for the stars),
and $n=4$ and \Reff\,$=6$\,kpc for the blue GCs.
The halo properties are $r_s = 48$\,kpc and $\rho_s = 2.3\times10^6 M_\odot$\,kpc$^{-3}$
($M_{200} = 4.6\times10^{12}M_\odot$ and $c_{200} = 7.1$).
The velocity dispersions are calculated using the Jeans equations \citep{Mamon05b}.
This $L^*$ model is also used for the $M_{\rm tot}(r)$ profile in Figure~\ref{fig:fracs}.
}
or dark matter halos with constant-density cores\footnote{The cored model is based on a logarithmic dark matter potential as discussed in
\citep{Thomas09}, with parameters $r_h=$\,5.0\,kpc and $v_h=$\,392\,km~s$^{-1}$ for the $4 L^*$ galaxy
(following the scaling relation for old ETGs in the Coma cluster),
and 3.5\,pc and 200\,km~s$^{-1}$ for the $L^*$ galaxy (similar to the young ETGs).},
as might arise either from baryonic effects or from alternative types of nonbaryonic dark matter (see discussion in \citealt{Newman13b}).

The top panel of Figure~\ref{fig:toy} illustrates some aspects of the mass distribution that may be learned through dynamical constraints
that extend from the centers of galaxies to their outer parts, as in SLUGGS.
Within the central few kpc of the galaxy, the circular velocity profile is hardly affected by large variations in the dark matter halo properties.
This is because the stars dominate the central regions, which is an advantage for constraining the stellar mass (and IMF)
using detailed data and models in these regions, but a disadvantage for learning about the  dark matter distribution
(e.g., \citealt{Treu10,Tortora10,Cappellari12}).

It is clearly important to employ data extending to larger radii to constrain the dark matter halos.
Here our SKiMS data have great potential, owing to their 2D coverage out to $\sim$\,3\,\Reff\ ($\sim$\,10\,kpc),
where the dark matter fraction is expected to be around 50\%.
These data, combined with central kinematics from SAURON/\atlas3d, may provide a unique opportunity to measure the slope of the
central dark matter profile -- which has so far been determined for dwarf, spiral, and brightest cluster galaxies
(e.g., \citealt{Oh11,Chemin11,Walker11,Newman13b}).

The GC kinematics data extend even farther out, to $\sim$\,8\,\Reff\ (tens of kpc), which
is close to the expected dark matter halo scale radius.
This is an important region to probe since it lies beyond the galaxy centers where myriad
baryonic effects could have altered the dark matter profiles.  At these large radii, it becomes possible to
test the primordial mass--concentration relation that is a robust prediction of
$\Lambda$CDM theory, at least for a given set of cosmological parameters (e.g., \citealt{Dutton14}).

While the SLUGGS stellar and GC kinematics datasets are each expected to provide powerful constraints on their own, 
the full potential of the survey will be realized when all of the data are integrated into ``global'' dynamical models
of the galaxies with large baselines in log-radius.   In particular, each galaxy has between two and four 
tracer populations (stars and one, two or three GC subpopulations, depending on how many distinct subpopulations are present) --
each of them providing independent constraints on the mass profile for a distinct radial regime.
This is an established approach for dwarf spheroidal galaxies with multiple stellar populations
(e.g., \citealt{Walker11,Amorisco13}),  and can now be applied to GC systems  with large datasets
(\citealt{Schuberth10,Napolitano14,Agnello14b}; Pota et al., in preparation).

Like GCs, PNe are also used as large-radius dynamical tracers.
Here the approach is somewhat simpler because the spatial distribution of PNe can be assumed
to follow the field starlight (see discussion in \citealt{Courteau14}).
However, GCs have the potent advantage of providing additional leverage through their subpopulations,
with the blue GCs in particular extending to larger radii than the PNe (and red GCs; see bottom panels of Figure~\ref{fig:toy}).
Given the extensive overlap between SLUGGS and PN-studied galaxies (see Section~\ref{sec:samp})
we plan to try incorporating joint GC and PN modeling constraints (cf.\ \citealt{Forbes12a}).

Most of our sample galaxies furthermore have detectable X-ray emitting hot gas halos (e.g., \citealt{OSullivan01,Boroson11}), 
some of which are suitable for mass analysis.
We have started to compare optically-based galaxy masses with those from X-rays and hence to test
the assumptions inherent in both methods. So far we have found in two galaxies that the 
mass estimates in the outer regions ($\sim$\,20--30\,kpc) as derived from the hot gas were much greater than those
from the GCs (and PNe; \citealt{Romanowsky09,Napolitano14}) -- which presumably reflects departures from
hydrostatic equilibrium.
More generally, the mass estimates from the SLUGGS galaxies should prove useful in understanding the properties
of the interstellar medium (e.g., \citealt{Humphrey13,Kim13}).

Given the above science goals and available data, there are a number of challenges for dynamical modeling. 
These include proper treatment of galaxy flattening,
permitting and constraining orbital anisotropy, and uncertainties in the stellar mass (from IMF variations in particular).
To tackle these issues, we are pursuing a layered modeling approach, ranging from simple mass estimators that may be applied
quickly to many galaxies \citep{Strader11,Agnello14a} to detailed, slower Schwarzschild-type orbit models 
\citep{Romanowsky01,Gebhardt09}.
We are also using techniques of intermediate complexity such as spherical Jeans models \citep{Romanowsky09,Napolitano14},
flattened Jeans models (\citealt{Cappellari08}, and in preparation), the projected virial theorem \citep{Agnello14b}, and
distribution function methods \citep{Deason12,Mamon13}.
It should be noted that many of these mass-modeling methods are designed to simultaneously
constrain the orbital anisotropy, e.g., by making use of higher-order information in the line-of-sight velocity distributions.

\section{Main Science Questions and Early Results}\label{sec:res}

Here we briefly summarize some highlights of the survey so far in the context of the scientific motivation outlined in 
Section~\ref{sec:intro}, and indicate the directions of our future work. A complete list of SLUGGS papers to date can be found at our survey
website\footnote{\tt http://sluggs.ucolick.org}.

\subsection{Evidence for Two-Phase Galaxy Assembly}\label{sec:twophase}

So far, the SLUGGS results support the dominance of a two-phase mode of galaxy assembly, as described in Section~\ref{sec:intro}.
The principal pieces of evidence include: broken radial metallicity gradients in the GC subpopulations \citep{Arnold11,Forbes11}, distinct core and halo kinematic behaviors of both the GCs and the underlying starlight \citep{Arnold11,Arnold14,Pota13a,Foster13},
and the presence of stellar streams, kinematical and positional fluctuations, and cold substructures in phase space 
(\citealt{Romanowsky12,Blom14};  see also Figure~\ref{fig:n4111} and Section~\ref{sec:GC kinematics}, as
well as references in Section~\ref{sec:intro} and other GC-based results in the literature from \citealt{Cote03,Schuberth10,Woodley11,DAbrusco14b}).  Rotation is found to be relatively weak in the outer regions,
at odds with general expectations for major mergers, but consonant with multiple, minor mergers in a cosmological context
\citep{Vitvitska02,Bekki08,RomanowskyFall12,Pota13a}.
Note that one rough prediction from cosmological accretion rates is that around half of nearby massive galaxies should
have residual substructure in their GC systems from minor mergers
(Figure~\ref{fig:fracs})\footnote{Here the calculation is based on the $L^*$ galaxy model from Section~\ref{sec:mass}
along with $\Lambda$CDM-based predictions for mergers with mass ratios of total $0.01$ and above at $z=0$ from \citet{Fakhouri10},
with no correction for galaxy Hubble type.
These total mass ratios correspond to stellar mass ratios of $\sim0.1$ \citep{Behroozi13} and hence
to considerable quantities of accreted stars and GCs.
The merger debris is assumed to persist as detectable substructure at all radii for 3 local dynamical times;
in reality, the debris would be preferentially deposited at larger radii.
}.

Despite this general agreement between observation and theory, there is a lingering puzzle about orbital anisotropy in the galaxy halos.
Almost any model of spheroidal galaxy formation (monolithic collapse, major mergers, multiple minor mergers)
predicts radially-biased orbits as a relic of the infall of gas and stars
(e.g., \citealt{Dekel05,McMillan07,Onorbe07,Sales07b,Prieto08}).
In the context of two-phase assembly, the radial bias should be prevalent especially in the most massive ellipticals,
where the ratio of accreted to in-situ stars and GCs is the highest \citep{Wu14}.
However, there are long-standing indications that the orbits of GCs around both the Milky Way and giant ellipticals are near-isotropic
or even tangentially-biased \citep{Frenk80,Romanowsky01,Cote01,Hwang08} -- 
a result which has only strengthened with the advent of data from SLUGGS
(\citealt{Romanowsky09,Deason12,Spitler12,Pota13a,Agnello14b}; Woodley et al., in preparation).

This observation has at times been attributed to a natural selection effect wherein GCs on near-radial orbits have been disrupted
from passing close to the galactic center.  However, this effect has not been shown to operate in the outer halos where the 
low anisotropy is found (cf.\ \citealt{Baumgardt98,Fall01,Vesperini03,Brockamp14}).
The only other explanaton so far proposed is preferential orbit dragging during galaxy formation by slow accretion and adiabatic contraction \citep{Agnello14b}.

\subsection{Distinct Subpopulations in GC Systems}\label{sec:bimodal}

For over two decades it has been realized that most massive galaxies possess GC systems with bimodal color distributions \citep{Zepf93,Ostrov93}.  Since the vast majority of these clusters have been shown to be old \citep[e.g.,][]{Strader05}, the color bimodality implies a corresponding bimodality in metallicity,
as is well-established for the Milky Way, where the metal-rich ``bulge'' and metal-poor ``halo'' clusters also have distinctly different kinematics and
spatial distributions \citep{Harris01}.
This observation is significant because GC formation is widely assumed to accompany the major star forming episodes in a galaxy's history, 
and therefore the presence of metallicity subpopulations suggests multiple epochs or phases of galaxy assembly that can be explored via their GC tracer populations (e.g., Section~\ref{sec:gcform}).

From a kinematic point of view, literature studies \citep[e.g.,][]{Cote03,Richtler04,Schuberth10,Lee10,Woodley10a} and more recently SLUGGS data have been used to establish the kinematic distinctiveness of
the subpopulations \citep{Strader11, Pota13a}, and to identify cases where more than the standard two subpopulations co-exist, due to a more complex or more recent accretion history \citep{Blom12a, Blom12b, Agnello14b}. Bimodality is also evident from the different ellipticities of the red and blue GC systems \citep{Park13}.

The GC bimodality paradigm has been disputed, in favor of a 
highly non-linear relation between metallicity and optical color that transforms a unimodal metallicity distribution into a bimodal color distribution \citep[e.g.,][]{Yoon06, Yoon11a, Yoon11b, Blakeslee12}.  
This scenario has been tested with near-infrared colors, which should be more linear, and in the few cases with adequate S/N, 
bimodality was confirmed \citep{Spitler08a,Chies12a,Cantiello14}.

\begin{figure}
\includegraphics[width=\columnwidth]{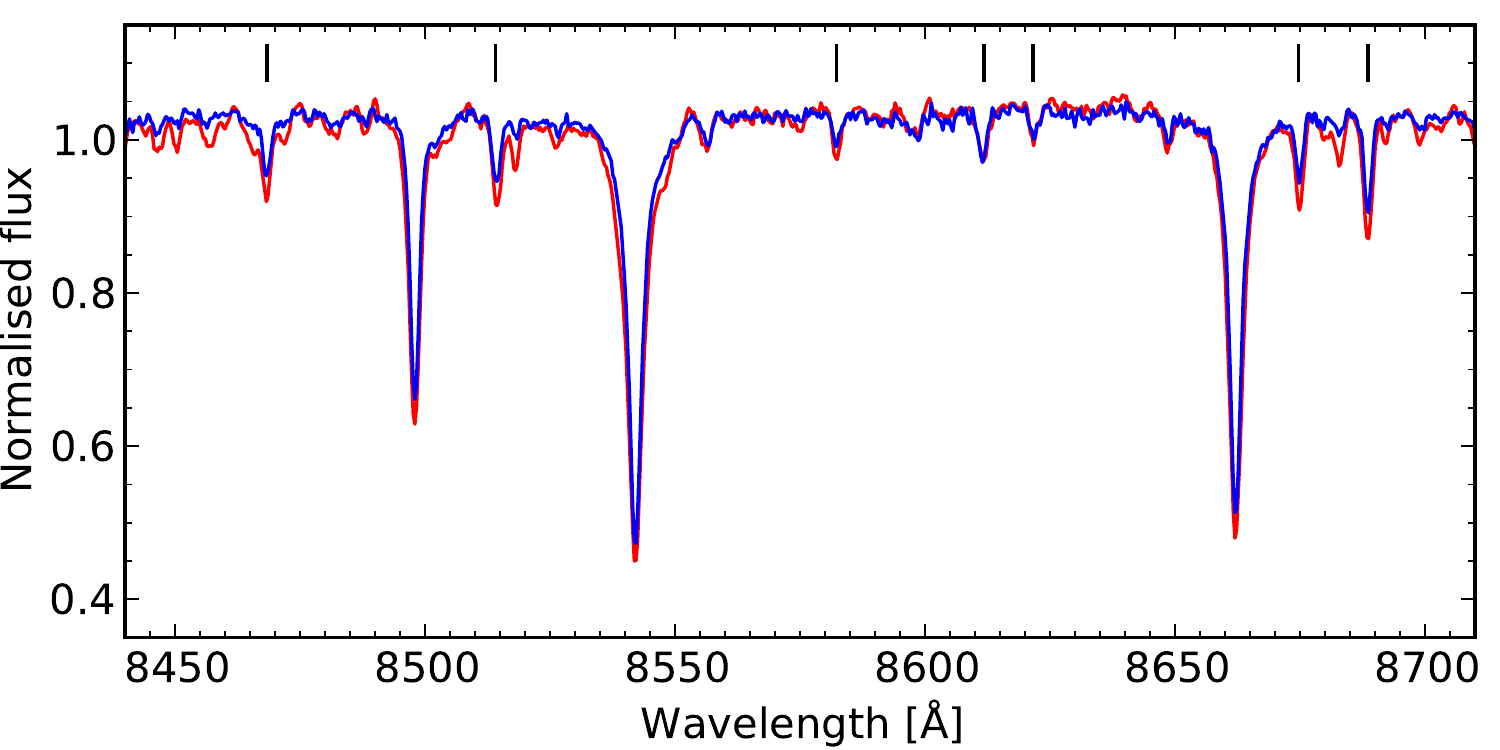}
\includegraphics[width=\columnwidth]{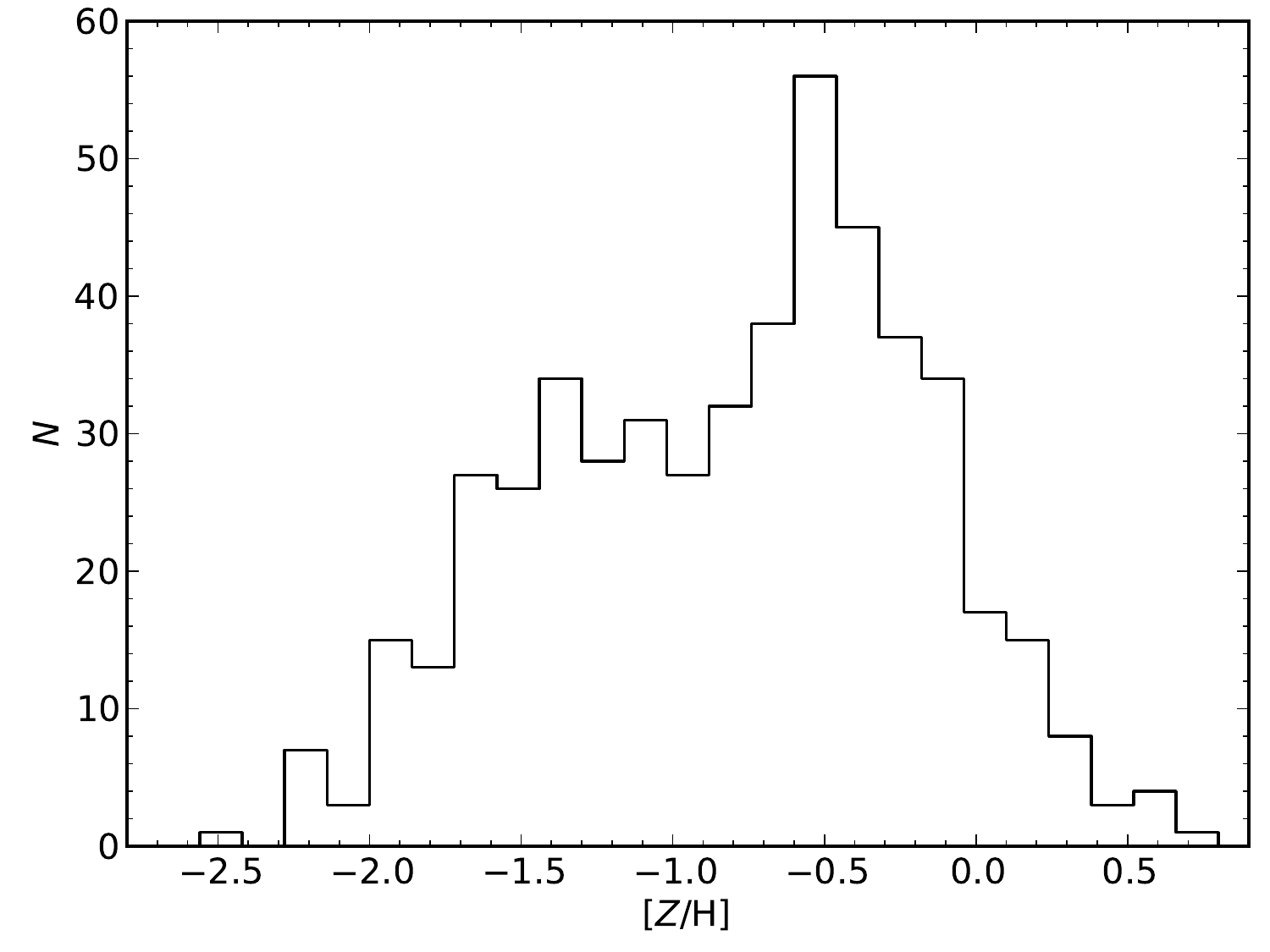}
\caption{Spectroscopic metallicities of GCs around SLUGGS galaxies.
The top panel shows stacks of the spectra around the CaT region for blue and red GC subsamples, with some of the iron lines marked.
The difference in metallicity between the two stacked spectra is apparent in the different CaT and iron line strengths.
The bottom panel shows the distribution of individual GC metallicities derived from the strength of the CaT, showing
502 GCs with [$\Delta Z$/H] $< 0.3$\,dex. Statistically, bimodality is strongly preferred over unimodality.
}
\label{fig:bimodal}
\end{figure}

The more definitive approach is to measure metallicities via spectroscopy, which has so far not been done with sufficiently large and
homogeneous GC data sets for drawing general conclusions (though see \citealt{Park12}).
We have now addressed this issue in a series of papers using SLUGGS data
\citep{Foster09,Foster11,Alves11, Brodie12, Usher12}.
We found that, in old GC populations, our CaT index is insensitive to alpha-element enhancement or horizontal branch morphology and can be reliably used to infer GC metallicities. 
Using CaT measurements for over 900 GCs in 11 galaxies, we established directly that the GC {\it metallicity} distributions were 
typically bimodal -- with some cases unclear owing to low-number statistics or to the likely presence of more than two metallicity subpopulations.

Here we note that, as with all aspects of GCs, it is important to exclude the highest-luminosity objects from the analysis
since they are heavily contaminated by ultra compact dwarfs -- which are likely to be chemically and dynamically distinct (Section \ref{sec:ucds}).
The natural focus on bright objects in earlier work may account for uncertain conclusions about bimodality
(e.g., \citealt{Cohen98}).
We have also found that the color--metallicity relation is a broken linear function that may vary with host galaxy mass \citep{Usher14}.

As a summary of these results, Figure~\ref{fig:bimodal} (top panel) shows ``stacked" GC spectra for 8 SLUGGS galaxies.
The blue stack contains 390 spectra from GCs with $0.72 < (g-i) < 0.88$\, has a S/N = 373~\AA$^{-1}$,
and has a CaT-based metallicity of [$Z$/H] = $-1.2$.
The red stack contains 353 spectra from GCs with $1.00 < (g-i) < 1.16$, has S/N = 419~\AA$^{-1}$,
and has a CaT-based metallicity of [$Z$/H] = $-0.4$.  
The two stacks differ not only in their CaT line strengths but also in the strengths of the interspersed, weaker iron lines (marked in the figure),
confirming that the CaT is a direct tracer of metallicity.
We are now further exploring the dependencies of other weak metal lines (such as Na, Mg, Ti) on metallicity, the IMF, and GC mass \citep{Usher14}.

The bottom panel of Figure~\ref{fig:bimodal} shows the spectroscopic metallicity distribution for all 502 GCs with high quality data 
($\Delta$[$Z$/H]$< 0.3$\,dex)
from a total of 11 galaxies from \citet{Usher12,Usher13}.
The individual galaxies show metallicity bimodality, which persists in the stacked distribution (no shifts were applied before stacking since
the metallicity peaks for each galaxy were all at similar locations).
Using the Gaussian mixture modeling code of \citet{Muratov10}, the $\chi^2$ fit, peak separation, and kurtosis of the distribution all strongly prefer bimodality to unimodality, with
$p < 0.001$, $p = 0.158$, and $p = 0.001$, respectively.
The code fitted peaks of [$Z$/H] = $-1.47 \pm 0.07$ and $-0.45 \pm 0.05$ dex.
Very similar results are obtained using the full sample of 908 GCs.

In summary, GC bimodality is supported by spectroscopic metallicities and the spatial distributions and kinematics of GC systems.
Although nonlinear relations between color and metallicity might in principle play a minor role in shaping GC color distributions, and the GC color/metallicity distributions of galaxies may occasionally not distinctly reveal the underlying GC subpopulations (e.g., due to a more complex accretion history),
the evidence strongly supports the result that the widely observed color bimodality is a true reflection of distinct subpopulations with inherently different mean metallicities.

\subsection{The Nature of Ultra-Compact Dwarfs}\label{sec:ucds}

Ultra-compact dwarfs (UCDs), compact spheroidal objects too large to fit within the general classification of GCs,  were 
predicted two decades ago to exist as the remnant nuclei of tidally stripped galaxies \citep{Bassino94},
and were subsequently confirmed by observations in the Virgo and Fornax clusters \citep{Hilker99, Drinkwater00}. 
Since that time there have been sustained observational efforts to understand whether these objects actually are stripped nuclei
(generally from dwarf ellipticals, but possibly also from compact ellipticals) or are extended star clusters  -- or both (e.g., \citealt{daRocha11,Norris11}).
Accumulating a large catalog of UCDs to begin studying them as a population is contingent on ascertaining their distances --
usually through spectroscopy.
SLUGGS thus provides a key opportunity to advance the UCD inventory by selecting suitable candidates as targets
along with normal GCs in our DEIMOS masks.

\begin{figure}
\includegraphics[width=\columnwidth]{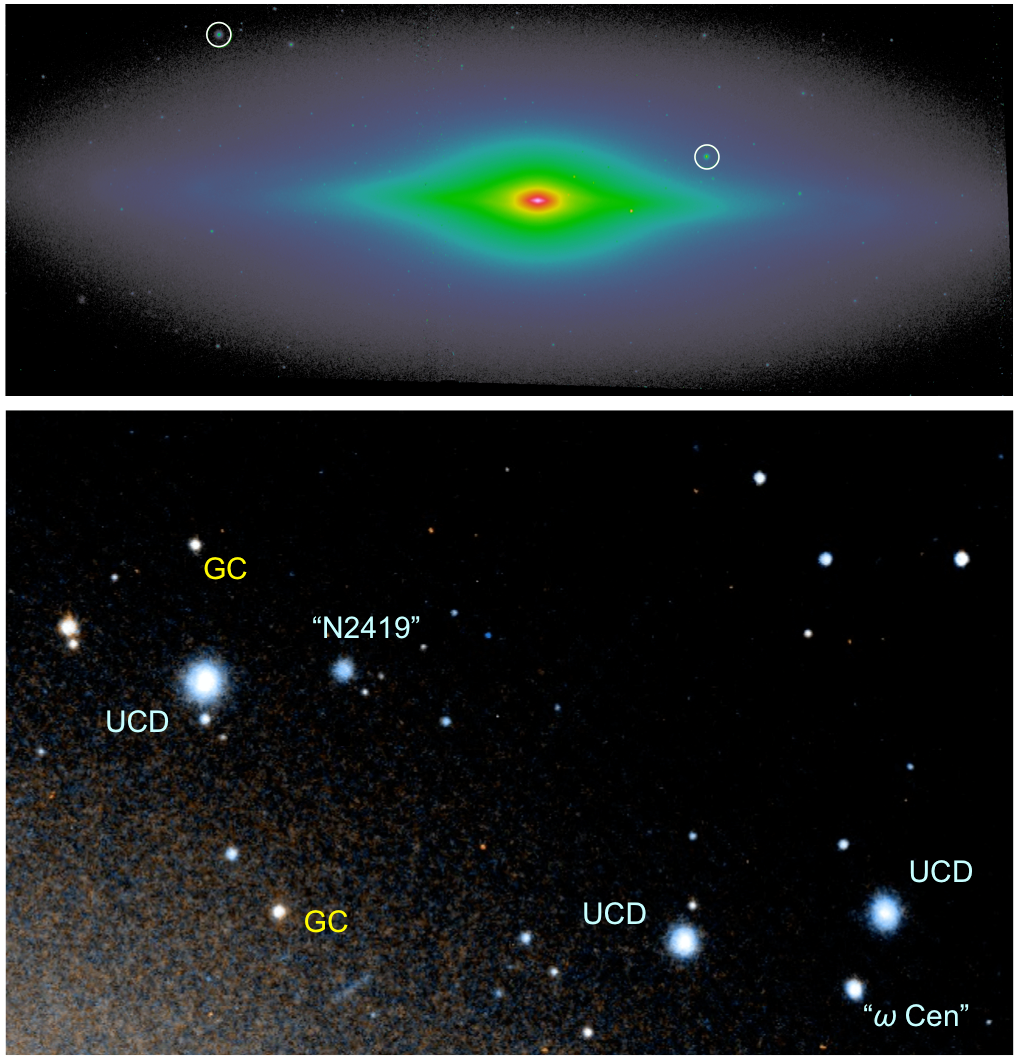}
\caption{Images of UCDs around SLUGGS galaxies, from {\it HST}/ACS.
Top panel: NGC~3115, a field lenticular galaxy, with two newly spectroscopically-confirmed UCDs marked with circles
(NGC~3115-UCD1 at upper left, and NGC~3115-AIMSS1 at right; \citealt{Jennings14,Norris14}).
Although such UCDs are bright enough to be seen in century-old photographs (e.g., \citealt{Pease17}),
their nature was only recognized through recent serendipitous spectroscopy.
Bottom panel: A zoomed-in region of the M87 inner halo, showing a variety of compact stellar systems, including GCs, an $\omega$\,Cen analog,
three ``classical'' UCDs, and a low-luminosity UCD analogous to NGC~2419 \citep{Brodie11}.
}
\label{fig:ucds}
\end{figure}

\begin{figure}
\includegraphics[width=\columnwidth]{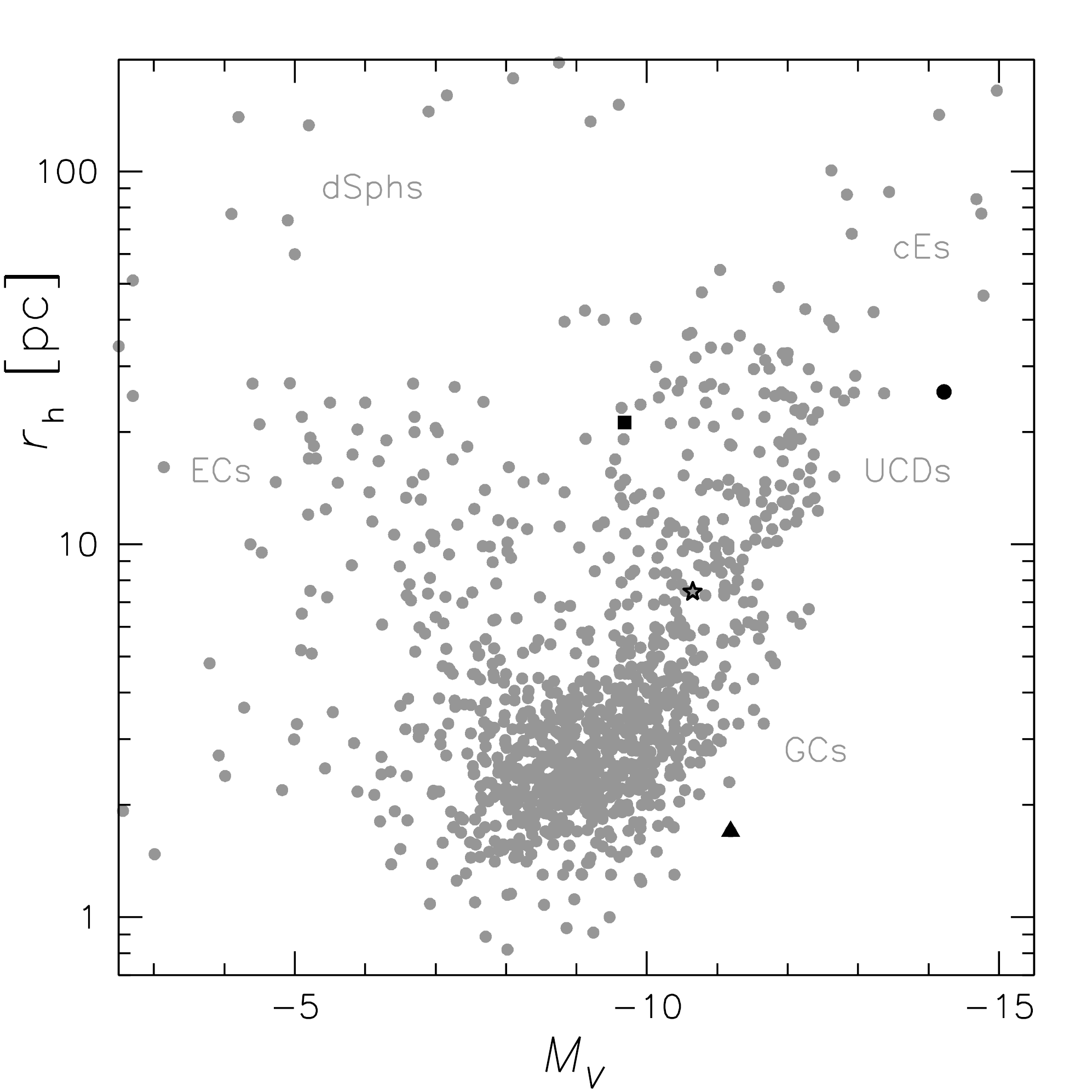}
\caption{Size--magnitude diagram of distance-confirmed compact stellar systems, showing the $V$-band absolute magnitude, and
the projected half-light radius. The data shown are a compilation begun in \citet{Brodie11}, with many more additions from the literature
(\citealt{Usher13,Forbes13,Norris14,Jennings14}), and includes $\sim$\,1070 sub-galactic 
objects\footnote{Electronic catalog available at {\tt http://sages.ucolick.org/spectral\_database.html}.}.
Schematic labels are provided  for different types of object (globular clusters, ultracompact dwarfs, extended clusters, dwarf spheroidals, compact ellipticals),
although we emphasize that the boundaries and distinctions between these separate populations are rather arbitrary.  It is a motivation of 
this work to better understand the properties of the various types of stellar systems and how they relate to one another in a physically motivated classification scheme.
Three objects of particular interest are highlighted by black symbols. $\omega$~Cen (star symbol) is the nearest example of a GC--UCD transition object.
NGC~2419 (square) is the nearest example of a low-luminosity UCD. N4494-UCD1 (triangle)
and M60-UCD1 (circle) are the densest known GC and galaxy, respectively.}
\label{fig:uber}
\end{figure}

Some results to date from SLUGGS and associated work include the first UCD  found around a non-cluster galaxy \citep{Hau09};  
the discovery of a new class of UCD, fainter than any UCDs previously studied \citep{Strader11,Brodie11}; 
the extension of objects into the so called ``zone of avoidance" in the size--luminosity plane for galaxies, UCDs and star clusters \citep{Forbes13}; as well as the discoveries of the densest known GC \citep{Foster11} and the densest known galaxy
\citep{Strader13}, which heralded a link between the UCD and cE populations 
(\citealt{Norris14}; see Figures~\ref{fig:ucds} and \ref{fig:uber}).

These results have eroded the classical distinction between star clusters and galaxies,  
since objects are now found to exist with a continuum of structural properties intermediate to these two populations.
Deciphering the evolutionary pathways becomes the next challenge after mapping out parameter space.
The UCD population appears to consist of both massive star clusters and stripped galaxies, 
with a ratio that changes across the size--luminosity diagram
(e.g., even objects with classical GC-like sizes of $\sim$\,4\,pc could originate from stripped nuclei; \citealt{Pfeffer13}).

\subsection{New Constraints on Globular Cluster Formation}\label{sec:gcform}

Given the central place of GCs in our efforts to understand the assembly of galaxies in the SLUGGS survey, it is worthwhile to revisit current thinking on the formation of GCs themselves. We take as our starting place the review of \citet{Brodie06}, while noting that the earlier reviews of \citet{Ashman98} 
and \citet{Harris01} still offer useful insights.

\citet{Brodie06} discussed the universal appearance of at least two subpopulations of GCs in all but the least massive galaxies, arguing that the metal-poor GCs likely formed at very high redshift ($z \sim 10$--15; \citealt{Elmegreen}), while the metal-rich GCs originated either in the dissipational merging that formed the main field star population of the host galaxy (at $z\ga2$ for $L^{*}$ ETGs; e.g., \citealt{Kruijssen12}), or/and in clumpy star formation regions in turbulent galaxies \citep{Kravtsov05,Shapiro10}. This scenario was consistent with the ages, metallicities, numbers, and spatial distributions of the GCs in each subpopulation (where the spatial distributions along with the kinematics are a reflection of the GC orbits).
{\it These are four fundamental observational constraints that must all be reproduced} for any model to be wholly successful in explaining the formation
of GC systems (and ultimately of the host galaxies themselves)\footnote{Other 
basic constraints more closely tied to the internal properties of GCs are 
the chemical abundance patterns, sizes, luminosity functions, and the correlations of these properties with the GC orbits.}.

Subsequent work has clarified and confirmed aspects of this picture while other parts have held up less well. 
As discussed in Section~\ref{sec:bimodal}, in-depth spectroscopic work has confirmed GC bimodality,
which has become a recognized benchmark for modeling the formation of GC systems.
As the general field of galaxy formation has shifted over the years from a focus on major mergers to a two-phase assembly
paradigm -- with early in-situ star formation and later satellite accretion -- the context and implications of GC modeling have shifted as well
(e.g., \citealt{Beasley02,Prieto08,Muratov10,Tonini13,Li14,Katz14}). 
Where it had been difficult to reproduce distinct GC metallicity peaks from major mergers, the
``new'' minor-merger dominated paradigm lends itself more naturally to bimodality,
with the metal-rich and metal-poor populations forming predominantly in the host and accreted galaxies, respectively.

However, while the models have now succeeded relatively well in reproducing GC metallicity distributions, they have to some
extent neglected the other critical constraints of ages and orbital distributions, which have also been firmed up over the years.
The compilations of spectroscopic age estimates  by \citet{Puzia05} and \citet{Strader05} have been
reinforced by subsequent findings of generally old GCs in ETGs
(e.g., \citealt{Pierce06a,Pierce06b,Cenarro07,Norris08,Woodley10b,Chies12b,Colucci13})\footnote{These results
are based almost exclusively on low-resolution optical spectra, where there is a long-standing degeneracy in
distinguishing young GCs from old ones with blue horizontal branches (e.g., \citealt{Burstein84,deFreitas95,Cohen03}).
This issue has now become very concrete with the dawning recognition that among massive GCs (the typical targets for
age studies), helium-enriched blue horizontal branches are probably very common (e.g., \citealt{Georgiev12,Milone14}). 
Hence occasional claims of more pervasive populations of young GCs -- without accounting for horizontal branch effects
as well as selection biases -- should be regarded as inconclusive \citep{Park12}.}.
However, there are also important novelties from studies of the Milky Way, whose GC system is assumed to be broadly analogous to those of ETGs.
Here, while isochrone fitting of GCs suggests ages ranging from $\sim$\,11 to 13.5 Gyr (formation epochs of $z \sim$\,2.5--20;
\citealt{Dotter11, Leaman13}), 
more precise ages are now claimed based on the white dwarf cooling sequence \citep{Hansen07,Hansen13,Bedin09}.
From studies of one metal-rich and two metal-poor Milky Way GCs, it is inferred that  they formed at $z\sim$\,2 and 3, respectively.
These conclusions fit in nicely with the existing picture for the metal-rich GCs -- but not for ``cosmological'' (very high $z$) formation
for the metal-poor GCs.

The small number statistics here are obviously problematic for drawing general conclusions about the overall Milky Way GC system, much less
the GCs of ETGs.
Moreover, since galaxies began forming at $z > 7.5$ \citep{Finkelstein13}, some 13~Gyr ago, it is likely that at least some metal-poor GCs also formed this early.
However, if most metal-poor GCs did share a $z\sim3$ formation epoch, then this would present a severe challenge to
any model that forms them mainly in association with low-mass dark matter subhalos or dwarf galaxies 
that are subsequently assembled into larger galaxies.
There is a well-established method for connecting the collapse epochs of subhalos with their spatial and orbital distributions at 
$z=0$, with the implication for metal-poor GCs that they must have formed very early, at $z\sim$\,9--11,
based on their tightly bound orbits
 \citep{Diemand05,Moore06,Spitler12,Corbett14}.
This issue has been noticed in cosmologically-motivated models that form GCs at later epochs, where the 
spatial distribution was found to be much too extended compared to observations 
(\citealt{Prieto08}; a similar problem is implied for the models of \citealt{Tonini13}).
 
Besides the ages issue, there is also the anisotropy puzzle discussed in Section~\ref{sec:twophase},
wherein the metal-poor GCs are not observed to follow radially-based orbits as expected in accretion scenarios.
The overall implication seems to be that a large fraction of these GCs formed in situ within the host galaxy, in a burst of metal-poor
star formation that is missing from current models.

This inference seems to mesh nicely with observational evidence in the Milky Way for
inner and outer components of the metal-poor stellar halo \citep{Carollo07},
which have been explained through cosmological hydrodynamical simulations 
as in-situ and accreted components \citep{Zolotov09,Font11a,Tissera14}.
A dual origin for the metal poor GCs is further supported by observed correlations between ages, metallicities, and orbits
\citep{Marin09,Forbes10,Keller12}, and between metallicity and galactocentric radius,
with steep central gradients that transition to constant mean metallicities in the outer parts --
both in the Milky Way and in early-type galaxies \citep{Harris01,Forbes11}. 

This metallicity gradient behavior has also been observed for metal-rich GCs \citep{Forbes11},
and taken together, the evidence from GCs suggests {\it three} phases of galaxy formation:
early in-situ metal-poor starbursts, later metal-rich central starbursts, and finally accretion of mixed-metallicity satellites.

The central metal-rich starburst phase may be described more specifically in the context of the current understanding
of massive galaxy formation, where the bulk of star formation does not occur in major mergers but in situ,
where cold streams from the cosmological web feed violently unstable disks at $z\sim2$ with high star formation rates
(e.g., \citealt{Parry09,Dekel09a,Hopkins10c}).
These disks are thought to host super-giant molecular clouds that are natural sites for GC formation
\citep{Shapiro10}, providing a tidy explanation for the origin of many metal-rich GCs.

The appeal of this ``wild-disk'' metal-rich GC formation scenario is enhanced by contrasting it with the more
traditional merger-formation scenario, as in \citet{Muratov10}.
Here most of the metal-rich GCs date from  $z < 1$, with very few from $z > 2$, in marked contrast with
the data on GC ages (\citealt{Dotter11}; but see also \citealt{Griffen10}, who were able to reproduce older
GCs through mergers).
The merger model does appear successful in at least reproducing the ``young'' branch of the Milky Way GC age--metallicity
relation, but such a relation is probably generic to scenarios of protracted GC formation epochs, e.g., within
dwarf galaxies that are later accreted\footnote{The latest merger-based models from \citet{Li14} appear to predict older
ages for metal-rich GCs around ETGs than the earlier work predicted for the Milky Way.
It remains to be seen how these models compare in detail with the GC age--metallicity relations both in the Milky Way
and in ETGs (accounting for different galaxy masses and environments, and also comparing to GC ages data
that are not specifically selected for youth).}. 

These developments from the last half-decade represent emerging insights into the mysterious origins of GCs, and of galaxies in general,
while the full picture still remains elusive.
We expect that substantial further progress will be made through considering the full chemodynamical constraints from the entire SLUGGS sample.

\section{Conclusions}\label{sec:conc}

We have here described the scientific motivation and technical details of the SLUGGS survey, whose unique strength is the collection of data to unprecedentedly wide fields of view around a statistically interesting set of galaxies covering a broad parameter base.  The data include two-dimensional maps of galaxy starlight kinematics and metallicities out to three effective radii, complemented and extended by imaging and spectroscopy of globular clusters out to eight effective radii.  

Many papers have already been published in connection with the survey and more are in active preparation. They address a wide variety of scientific questions that ultimately relate to our understanding of the star formation and assembly histories of galaxies.   The brief overview of selected results given above reinforces the validity of the original drivers for the work:  the inner regions of galaxies are poor predictors of the behavior of the outer regions, 
which furthermore contain the relics, albeit often well-hidden in the positional--kinematic--metallicity phase-space, of the formation history of the entire galaxy. 

Although the sampled number of galaxies is modest, the particular strengths of SLUGGS are its superior velocity resolution and radial extent over any existing 2D chemodynamical survey of early-type galaxies.  These advantages will be important for interpreting data from future large surveys (e.g., MaNGA; Section~\ref{sec:survcomp}); the synergy between wide radial extent for representative galaxies and large statistical samples promises robust insights into the assembly histories of galaxies over a wide range of masses and environments.

\acknowledgments

We are grateful to Beth Johnson for assisting with SDSS images, Steve Romanowsky for help with the logo,
and Tomer Tal for kindly allowing us to use his image of NGC 4111. 
We thank  Michele Cappellari, Jesus Falc\'on-Barroso, Jenny Greene, Thorsten Naab, and the referee for helpful comments.
Data presented herein were obtained at the W.~M.~Keck Observatory, which is operated as a scientific partnership between Caltech, UC, 
and NASA, and at Subaru (operated by National Astronomical Observatory of Japan) via Gemini (GN-2006B-C-18, GN-2008A-C-12) and Keck time exchanges. Some of the data were acquired through SMOKA, which is operated by the Astronomy Data Center, National Astronomical Observatory of Japan. This work was supported by the NSF through grants AST-0909237, AST-1109878, and AST-1211995. DAF thanks the ARC for financial support via DP130100388. LS was supported by the ARC Discovery Program grant DP0770233.  This research made use of Montage, funded by the National Aeronautics and Space Administration's Earth Science Technology Office, Computation Technologies Project, under Cooperative Agreement number NCC5-626 between NASA and the California Institute of Technology. Montage is maintained by the NASA/IPAC Infrared Science Archive.\\

\bibliographystyle{apj}
\bibliography{biblio}

\appendix\label{sec:app}

Here we describe some further details of the galaxy sample parameters, as discussed in Section~\ref{sec:samp} and adopted in 
Figure~\ref{fig:params0} and Table~\ref{info}.
\\

\begin{itemize}

\item
$M_K$: The $K$-band absolute magnitudes are derived from the 2MASS extended source catalog \citep{Jarrett00},
using the extrapolated magnitude parameter {\tt k\_m\_ext}.
This magnitude is extinction-corrected using the procedure described in \citet{Cappellari11a}.
An exception is NGC~2974, whose 2MASS photometry is severely impacted by a bright foreground star, and 
instead we adopt the $M_K$ value from \atlas3d\ \citep{Cappellari11a}, who used their own $I$-band photometry \citep{Cappellari06} 
and an assumed $I-K$ color to correct to $K$-band (M. Cappellari, priv.\ communication; NGC~4486A was the other \atlas3d\ galaxy
similarly affected and corrected).
See Section~\ref{sec:samp} for further caveats about the 2MASS magnitudes.

\item
$D$: For the distances, we adopt estimates based on surface-brightness fluctuations, using the {\it HST}-based work of \citet{Blakeslee09} for galaxies
in the vicinity of Virgo, and the ground-based work of \citet{Tonry01} otherwise, after subtracting a distance modulus of $-0.06$\,mag.
Note that \atlas3d\ used \citet{Mei07} for the Virgo galaxies, while we have followed the recalibration of \citet{Blakeslee09}.
For most galaxies, this makes a difference of less than 0.1\,Mpc (1\% in distance), but for the brightest galaxies such as M87, the changes are up to 
$\sim$\,1\,Mpc ($\sim 5$\% in distance, $\sim 10$\% in luminosity).
We have also estimated a distance to the Virgo A sub-cluster (centered on M87), following \cite{Mei07}
by averaging the distances of all 32 galaxies within 2\,deg of M87, and finding $16.7 \pm 0.2$\,Mpc --
which is nicely consistent with the $16.7 \pm 0.6$\,Mpc distance to M87 itself, and to 
other results on the M87 distance \citep{Bird10}.

NGC~1400 and NGC~1407 in the Eridanus~A group are another case of special concern, as a wide range of distances to
these galaxies has been used in the literature.
The very high relative velocities of these two galaxies suggest that they are very close in line-of-sight distance, so we use the same,
averaged distance of 26.8\, Mpc for both galaxies, from \citet{Tonry01}.
Note that a further adjustment for data quality \citep{Blakeslee10b} would imply a distance of 26.5\,Mpc, while
an {\it HST}-based distance to NGC~1407 from \citet{Cantiello05}, after recalibration to the distance to NGC~1344 in \citet{Blakeslee09},
gives 26.3\,Mpc.  However, for the sake of homogeneity with the other galaxies in our sample, we do not apply these minor corrections.

\item
$\sigma$: For the central stellar velocity dispersion, we use the value within 1\,kpc from \atlas3d, $\sigma_{\rm kpc}$  \citep{Cappellari13a}.
For the non-\atlas3d\ galaxies, we use the central velocity dispersion {\tt vdis} from HyperLeda,
applying the calibration reported by \atlas3d, 
$\sigma_{\rm kpc} \simeq  (\sigma/0.55 \, {\rm km\,s}^{-1})^{0.892}$.
Note that this calibration deviates systematically for the highest-$\sigma$ galaxies, perhaps owing to template issues with
alpha-element enhanced stellar populations, but this issue is not significant for the purposes of sample selection.

\item
$V/\sigma$: To characterize the rotation, we use $V/\sigma_{\rm e/2}$ from \atlas3d\ \citep{Emsellem11}, which is
the aperture-averaged value of rotation divided by dispersion with 0.5\,\Reff.
Galaxies with low and high degrees of rotational support have $V/\sigma_{\rm e/2} \sim 0$ and $\sim 1$, respectively.
We do {\it not} use the value within 1\,\Reff\ as reported by \atlas3d, since the SAURON kinematics measurements did not actually
extend out to 1\,\Reff\ for many galaxies, and in particular not for galaxies in the $L^*$ and brighter luminosity range
that is the focus of SLUGGS (cf.\ figure~1 in \citealt{Arnold14}).
For the non-\atlas3d\ galaxies, 
we use an approximate calibration for $V/\sigma$
(equation~23 of \citealt{Cappellari07})
based on long-slit data (e.g., \citealt{vanderMarel94a}).

\item
\Reff:
The effective radii are taken from a calibrated average of 2MASS and RC3 results \citep{RC3}, 
as discussed in \citet{Cappellari13a}.
Note that these values are adopted for the sake of uniformity and not for accuracy.
For the brightest galaxies in particular (NGC~4365, NGC~4374, NGC~4486, NGC~4649),
the effective radii derived through deeper optical imaging are twice as large \citep{Kormendy09},
which should be taken into account for more detailed studies, e.g., dynamical modeling.

\item
Morph., $T_{\rm Hub}$: 
The galaxy morphologies (E, S0, etc.)~are taken from NED, combining the RC3 and RSA classifications
(since these sometimes disagree).
The Hubble stage parameter $T_{\rm Hub}$ is taken from the morphological type code of HyperLeda \citep{Paturel03}.
The exception is NGC~2974, for which we add a partial S0 classification since it hosts spiral arms and 
was classified as an S0 or Sa by all six observers (including G.\ de Vaucouleurs) in \cite{Naim95}.
We also assign it the numerical value of $T_{\rm Hub} = -2.1$.

\item
P.A., $\epsilon$:
The photometric position angles and ellipticities are taken from \citet{Krajnovic11} for the \atlas3d\ galaxies,
where it represents the large-scale values in the $r$-band, at typically 2.5--3\,\Reff.
The alternative values of $\epsilon_{\rm e}$ (average within 1\,\Reff) used elsewhere in \atlas3d\ do not capture
well the overall shapes of the galaxies, e.g., $\epsilon_{\rm e} = 0.36$ for the near edge-on lenticular, NGC~1023,
as compared to $0.63$ from the outer isophotes.
For the non-\atlas3d\ galaxies, we take the values from the RC3.

\item
$a_4/a$: 
This parameter represents the isophote shape, in percent deviation from a perfect ellipse.
Negative values trace boxiness, and positive values trace diskiness.
For the \atlas3d\ galaxies, the values as used in \citet{Emsellem11} are taken from the survey webpage\footnote{\tt http://www-astro.physics.ox.ac.uk/atlas3d/}.
Note that these values are weighted averages within 1\,\Reff\ and do not necessarily represent the overall galaxy properties.
For the non-\atlas3d\ galaxies, approximate values for the central regions are obtained from the CGS survey \citep{Li11}.

\item
$\gamma^\prime$ :
The central logarithmic density negative slope of the stellar surface brightness profile is taken for the \atlas3d\ galaxies from
\citet{Krajnovic13b}.  Cored galaxies have $\gamma^\prime < 0.3$, cusped galaxies have $\gamma^\prime > 0.5$, and other galaxies are intermediate.
For the non-\atlas3d\ galaxies, we draw from \citet{Lauer07}, \citet{Spolaor08a}, and \citet{Rusli13}.

\item
$V_{\rm sys}$:
The systemic velocity is taken from \atlas3d\ \citep{Cappellari11a}, and otherwise from NED.

\item
$\rho_{\rm env}$, Env.:
The environmental density is expressed in number of galaxies per Mpc$^{3}$ and
is taken from the \citet{Tully88} Nearby Galaxies Catalogue
(this parameter is available for the full SLUGGS sample and for 59\% of the \atlas3d\ sample).
The densities for NGC~1400 and NGC~1407 arguably ought to be higher, with the high relative velocity
relecting a recent infall of NGC~1400 rather than a large physical separation
(see \citealt{Romanowsky09,Su14}).
To address this issue, we have given NGC~1400 the same $\rho_{\rm env}$ as NGC~1407.
The galaxies are also classified as cluster, group, or field members according to their group assignments by Tully,
where we assign galaxies to the field if there is no other group member with at least one-third its luminosity.
The typical $\log \rho_{\rm env}$ values for these categories are $0.5$, $-0.3$, and $-1$.
The ``Virgo'' galaxy NGC~4365 is assigned to a background group, based on SBF distance measurement.

\end{itemize}

\begin{figure*}
\includegraphics[width=\textwidth]{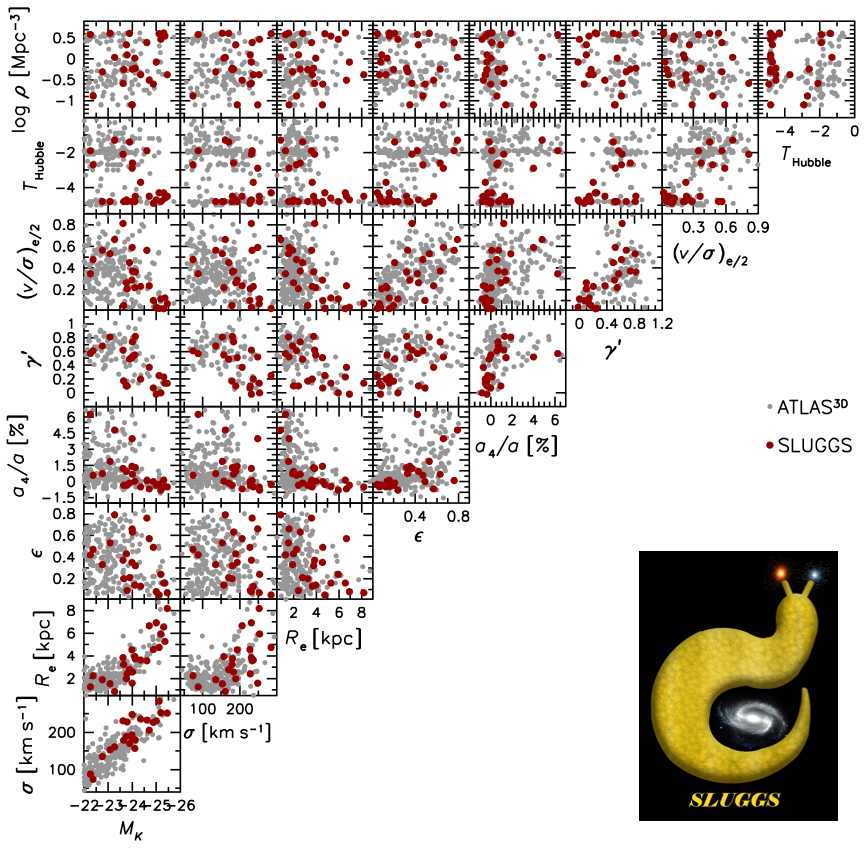}
\caption{Multi-dimensional parameter space of early-type galaxies in the nearby universe.
Red points are from the SLUGGS survey; smaller gray points are from \atlas3d.
In broad terms, three main classes of galaxies may be recognized (cf.\ \citealt{Kormendy96}), 
with the associated SLUGGS galaxies as follows:
boxy, round, massive, centrally slowly rotating ellipticals with central stellar cores:
NGCs 720, 1407, 3608, 4365, 4374, 4486, 4649, 5846;
disky, flattened, less massive, centrally fast-rotating ellipticals with central stellar cusps:
NGCs 821, 1400, 2768, 3377, 4278, 4473, 4494, 4564, 4697;
lenticulars:
NGCs 1023, 2974, 3115, 4111, 4459, 4474, 4526, 7457.
In spite of the limited pool of galaxy candidates in the very nearby universe ($D \lsim$ 25\,Mpc), our sample provides excellent coverage of galaxy parameter space (see main text for detailed description).
 }\label{fig:params}
\end{figure*}

The full grid of parameter space is shown in Figure~\ref{fig:params}, with the SLUGGS sample compared to the \atlas3d\
inventory of galaxies in the nearby universe.  Note that the distances (and therefore the magnitudes) of the \atlas3d\
galaxies in Virgo have been adjusted so as to be consistent with the SLUGGS adopted distances,
and the $T_{\rm Hub}$ values of all the \atlas3d\ galaxies have likewise been updated.

\end{document}